\def\year{2022}\relax
\documentclass[letterpaper]{article} 
\usepackage{aaai22}  
\usepackage{times}  
\usepackage{helvet}  
\usepackage{courier}  
\usepackage[hyphens]{url}  
\usepackage{graphicx} 
\usepackage{natbib}  
\usepackage{caption} 
\DeclareCaptionStyle{ruled}{labelfont=normalfont,labelsep=colon,strut=off} 
\usepackage{threeparttable}
\usepackage{booktabs}
\usepackage{makecell}
\usepackage{amssymb}
\usepackage{tikz}
\usepackage{algorithm}
\usepackage{algorithmic}
\usepackage{pgfplots}
\usepackage{subfigure}
\usetikzlibrary{positioning}
\usetikzlibrary{calc}
\usetikzlibrary{arrows}
\usetikzlibrary{backgrounds}
\usepackage{xcolor}
\definecolor{ublue}{rgb}{0.152,0.250,0.545}
\definecolor{ugreen}{rgb}{0,0.5,0}
\definecolor{lgreen}{rgb}{0.9,1,0.8}
\definecolor{amber}{rgb}{1.0, 0.75, 0.0}
\definecolor{xtgreen}{rgb}{0.914,0.945,0.902}
\definecolor{lightgray}{gray}{0.85}
\usepackage{newfloat}
\usepackage{listings}
\floatstyle{ruled}
\newfloat{listing}{tb}{lst}{}
\floatname{listing}{Listing}
\title{Improving End-to-end Speech Translation by Leveraging Auxiliary\\Speech and Text Data}
\author{
    Yuhao Zhang\textsuperscript{\rm 1},
    Chen Xu\textsuperscript{\rm 1},
    Bojie Hu\textsuperscript{\rm 3},
    Chunliang Zhang\textsuperscript{\rm 1,2},
    Tong Xiao\textsuperscript{\rm 1,2}\footnote{Corresponding author.},
    Jingbo Zhu \textsuperscript{\rm 1,2}
}
\affiliations{
    \textsuperscript{\rm 1} School of Computer Science and Engineering, Northeastern University, Shenyang, China \\
    \textsuperscript{\rm 2} NiuTrans Research, Shenyang, China \\
    \textsuperscript{\rm 3} Tencent Minority-Mandarin Translation, Beijing, China\\
    yoohao.zhang@gmail.com, xuchenneu@outlook.com, bojiehu@tencent.com, \\ \{zhangchunliang, xiaotong, zhujingbo\}@mail.neu.edu.cn

}

\begin{document}

\maketitle

\begin{abstract}
We present a method for introducing a text encoder into pre-trained end-to-end speech translation systems. It enhances the ability of adapting one modality (i.e., source-language speech) to another (i.e., source-language text). Thus, the speech translation model can learn from both unlabeled and labeled data, especially when the source-language text data is abundant. Beyond this, we present a denoising method to build a robust text encoder that can deal with both normal and noisy text data. Our system sets new state-of-the-arts on the MuST-C En-De, En-Fr, and LibriSpeech En-Fr tasks.
\end{abstract}

\section{Introduction}
In Speech Translation (ST), End-to-End (E2E) neural approaches have gained attraction as a promising line of research towards systems with lower latency and less error propagation. However, developing models of this type can be challenging because the aligned speech-to-translation data is scarce \cite{wang2020bridging,dong2021listen,pmlr-v139-zheng21a,tang-etal-2021-improving}. This leads researchers to explore methods that resort to large-scale unlabeled data. A simple one is to use pre-trained models to encode acoustic and/or textual input \cite{pino20_interspeech,ye2021end}, whereas others train ST models using additional data of either Automatic Speech Recognition (ASR) or Machine Translation (MT), or both \cite{wang-etal-2020-curriculum,xu-etal-2021-stacked,9414703}.

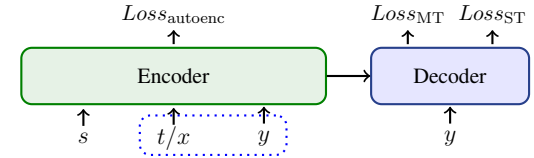
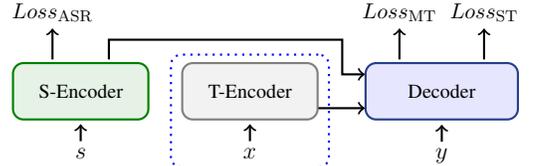
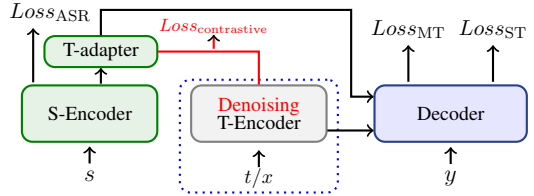
\begin{figure}[t!]

\centering

\subfigure[\mbox {1 universal encoder \cite{pmlr-v139-zheng21a}}]
{
\tikzstyle{encoder} = [rectangle,thick,rounded corners,minimum width=5.4cm,minimum height=1cm,text centered,draw=ugreen,fill=ugreen!10,inner sep=0.1cm]
\tikzstyle{decoder} = [rectangle,thick,rounded corners,minimum width=2.8cm,minimum height=1cm,text centered,draw=ublue,fill=blue!10,inner sep=0.1cm]
\tikzstyle{frame} = [rectangle,thick,dotted, rounded corners,minimum width=2.6cm,minimum height=0.7cm,text centered,draw=blue]

\begin{tikzpicture}[node distance = 0,scale = 0.75]
\tikzstyle{every node}=[scale=0.75]
\node(encoder)[encoder]{\large{Encoder}};
\node(frame)[frame,below of = encoder,xshift=0.7cm,yshift=-1.05cm]{};
\node(decoder)[decoder,right of = encoder,xshift=4.9cm,yshift=0cm]{\large{Decoder}};
\node(emptyword)[anchor=east] at (encoder.west) {\color{white} x};

\node(input_s)[below of = encoder,xshift=-1.6cm,yshift=-1.1cm]{\Large{$s$}};
\node(input_tx)[below of = encoder,xshift=0cm,yshift=-1.1cm]{\Large{$t/x$}};
\node(input_y)[below of = encoder,xshift=1.6cm,yshift=-1.1cm]{\Large{$y$}};
\node(input_de)[below of = decoder,xshift=0cm,yshift=-1.1cm]{\Large{$y$}};

\draw[->,thick](encoder.east)to(decoder.west);
\draw[->,thick](input_s.north)to([yshift=0.3cm]input_s.north);
\draw[->,thick]([yshift=-0.1cm]input_tx.north)to([yshift=0.2cm]input_tx.north);
\draw[->,thick](input_y.north)to([yshift=0.3cm]input_y.north);
\draw[->,thick](input_de.north)to([yshift=0.3cm]input_de.north);

\node(output_loss1)[above of = encoder,xshift=0cm,yshift=1.1cm]{\large{$Loss_{\mathrm{autoenc}}$}};
\node(output_loss2)[above of = decoder,xshift=-0.75cm,yshift=1.1cm]{\large{$Loss_{\mathrm{MT}}$}};
\node(output_loss3)[above of = decoder,xshift=0.75cm,yshift=1.1cm]{\large{$Loss_{\mathrm{ST}}$}};

\draw[->,thick]([yshift=-0.25cm]output_loss1.south)to(output_loss1.south);
\draw[->,thick]([yshift=-0.25cm]output_loss2.south)to(output_loss2.south);
\draw[->,thick]([yshift=-0.25cm]output_loss3.south)to(output_loss3.south);

\end{tikzpicture}
}
\\

\vspace{-0.5em}
\subfigure[\mbox {2 individual encoders \cite{li-etal-2021-multilingual}}]{
\tikzstyle{encoder} = [rectangle,thick,rounded corners,minimum width=2.4cm,minimum height=1cm,text centered,draw=ugreen,fill=ugreen!10,align=center,inner sep=0.1cm]
\tikzstyle{decoder} = [rectangle,thick,rounded corners,minimum width=2.7cm,minimum height=1cm,text centered,draw=ublue,fill=blue!10,align=center,inner sep=0.1cm]
\tikzstyle{frame} = [rectangle,thick,dotted, rounded corners,minimum width=2.8cm,minimum height=2cm,text centered,draw=blue]

\begin{tikzpicture}[node distance = 0,scale = 0.75]
\tikzstyle{every node}=[scale=0.75]
\node(encoder_s)[encoder]{S-Encoder};
\node(encoder_t)[encoder,right of=encoder_s,draw=gray,fill=gray!10,xshift=3cm,yshift=0cm]{T-Encoder};
\node(frame)[frame,below of = encoder_t,xshift=0cm,yshift=-0.35cm]{};
\node(decoder)[decoder,right of = encoder_t,xshift=3.4cm,yshift=0cm]{Decoder};

\node(input_s)[below of = encoder_s,xshift=0cm,yshift=-1.1cm]{\Large{$s$}};
\node(input_tx)[below of = encoder_t,xshift=0cm,yshift=-1.1cm]{\Large{$x$}};
\node(input_de)[below of = decoder,xshift=0cm,yshift=-1.15cm]{\Large{$y$}};

\draw[->,thick]([yshift=-0.3cm]encoder_t.east)to([yshift=-0.3cm]decoder.west);
\draw[->,thick](input_s.north)to([yshift=0.3cm]input_s.north);
\draw[->,thick](input_tx.north)to([yshift=0.3cm]input_tx.north);
\draw[->,thick](input_de.north)to([yshift=0.3cm]input_de.north);

\node(output_loss1)[above of = encoder_s,xshift=-0.5cm,yshift=1.4cm]{\large{$Loss_{\mathrm{ASR}}$}};
\node(output_loss2)[above of = decoder,xshift=-0.75cm,yshift=1.4cm]{\large{$Loss_{\mathrm{MT}}$}};
\node(output_loss3)[above of = decoder,xshift=0.75cm,yshift=1.4cm]{\large{$Loss_{\mathrm{ST}}$}};

\draw[->,thick]([yshift=-0.55cm]output_loss1.south)to(output_loss1.south);
\draw[->,thick]([yshift=-0.55cm]output_loss2.south)to(output_loss2.south);
\draw[->,thick]([yshift=-0.55cm]output_loss3.south)to(output_loss3.south);
\draw[->,thick]([xshift=1cm,yshift=-0.55cm]output_loss1.south)to([xshift=1cm,yshift=-0.2cm]output_loss1.south)
-|([xshift=-0.4cm,yshift=0.3cm]decoder.west)to([yshift=0.3cm]decoder.west);

\end{tikzpicture}
}
\\

\vspace{-0.5em}
\subfigure[\mbox {2 correlated encoders + denoising (this work)}]
{
\tikzstyle{encoder} = [rectangle,thick,rounded corners,minimum width=2.4cm,minimum height=1cm,text centered,draw=ugreen,fill=ugreen!10,align=center]
\tikzstyle{adapter} = [rectangle,thick,rounded corners,minimum width=2.0cm,minimum height=0.5cm,text centered,draw=ugreen,fill=ugreen!10,align=center]
\tikzstyle{decoder} = [rectangle,thick,rounded corners,minimum width=2.7cm,minimum height=1cm,text centered,draw=ublue,fill=blue!10,align=center]
\tikzstyle{frame} = [rectangle,thick,dotted, rounded corners,minimum width=2.8cm,minimum height=2cm,text centered,draw=blue]

\begin{tikzpicture}[node distance = 0,scale = 0.75]
\tikzstyle{every node}=[scale=0.75]
\node(encoder_s)[encoder]{S-Encoder};
\node(t_adapter)[adapter,above of=encoder_s,yshift=1.1cm,xshift=0.2cm]{T-adapter};
\node(encoder_t)[encoder,right of=encoder_s,draw=gray,fill=gray!10,xshift=3cm,yshift=0cm]{ \textcolor{red}{Denoising}\\[-0.5ex] T-Encoder};
\node(frame)[frame,below of = encoder_t,xshift=0cm,yshift=-0.4cm]{};
\node(decoder)[decoder,right of = encoder_t,xshift=3.4cm,yshift=0cm]{Decoder};

\node(input_s)[below of = encoder_s,xshift=0cm,yshift=-1.1cm]{\Large{$s$}};
\node(input_tx)[below of = encoder_t,xshift=0cm,yshift=-1.15cm]{$t/x$};
\node(input_de)[below of = decoder,xshift=0cm,yshift=-1.15cm]{\Large{$y$}};

\draw[->,thick]([yshift=-0.28cm]t_adapter.south)to(t_adapter.south);
\draw[->,thick]([yshift=-0.3cm]encoder_t.east)to([yshift=-0.3cm]decoder.west);
\draw[->,thick](input_s.north)to([yshift=0.3cm]input_s.north);
\draw[->,thick]([yshift=-0.1cm]input_tx.north)to([yshift=0.2cm]input_tx.north);
\draw[->,thick](input_de.north)to([yshift=0.3cm]input_de.north);

\node(output_loss1)[above of = encoder_s,xshift=-0.7cm,yshift=1.8cm]{\large{$Loss_{\mathrm{ASR}}$}};
\node(output_loss2)[above of = decoder,xshift=-0.75cm,yshift=1.5cm]{\large{$Loss_{\mathrm{MT}}$}};
\node(output_loss3)[above of = decoder,xshift=0.75cm,yshift=1.5cm]{\large{$Loss_{\mathrm{ST}}$}};
\node(output_loss4)[right of = t_adapter, xshift=2cm,yshift=0.45cm,color=red]{\small{$Loss_{\mathrm{contrastive}}$}};

\draw[->,thick]([xshift=-0.3cm,yshift=-0.96cm]output_loss1.south)to([xshift=-0.3cm]output_loss1.south);
\draw[->,thick]([yshift=-0.65cm]output_loss2.south)to(output_loss2.south);
\draw[->,thick]([yshift=-0.65cm]output_loss3.south)to(output_loss3.south);
\draw[->,thick](t_adapter.north)to([yshift=0.45cm]t_adapter.north)
-|([xshift=-0.4cm,yshift=0.3cm]decoder.west)to([yshift=0.3cm]decoder.west);
\draw[-,thick,draw=red]([xshift=-0cm,yshift=0cm]t_adapter.east) --(t_adapter.east-|encoder_t.north)-- (encoder_t.north);

\draw[->,thick,draw=red]([yshift=-0.15cm]output_loss4.south)to([yshift=0.1cm]output_loss4.south);
\end{tikzpicture}
}

\vspace{-0.5em}
\caption{Model architectures. Dotted boxes mean that the items are dropped in ST tuning and inference. $s=$ acoustic signal sequence, $t =$ transcription, $x=$ source-language word sequence, and $y=$ target-language word sequence.}
\label{fig:model-types}
\vspace{-0.5em}

\end{figure}

Such a paradigm provides an opportunity to make use of both labeled and unlabeled data, say, the speech, text, speech-to-transcription, text-to-text, and speech-to-text data. For example, one can feed all available data into an autoencoder to train ST models \cite{pmlr-v139-zheng21a}. More recently, it has been found that stronger results can be achieved by using an explicit text encoder to ease the training on the MT data \cite{li-etal-2021-multilingual}.

\begin{figure*}[t]
\centering

\definecolor{Tan}{rgb}{0.8,0.61,0.31}
\definecolor{LightCoral}{rgb}{0.94117,0.50196,0.50196}
\definecolor{color1}{rgb}{0.7,0.5,0.65}

\tikzstyle{input} = [rectangle,thick,rounded corners,minimum width=4.8cm,minimum height=0.7cm,text centered,draw=Tan,fill=Tan!10,align=center]
\tikzstyle{encoder} = [rectangle,thick,rounded corners,minimum width=2.8cm,minimum height=1.2cm,text centered,draw=ugreen,fill=ugreen!10,align=center,font=\small]

\tikzstyle{decoder} = [rectangle,thick,rounded corners,minimum width=2.8cm,minimum height=2.8cm,text centered,draw=ublue,fill=blue!10,align=center]
\tikzstyle{num} = [circle,minimum size=0.4cm,text centered,fill=blue!50,align=center,inner sep=0.08cm,font=\small,text=white]

\begin{tikzpicture}[node distance = 0,scale = 0.85]
\begin{scope}[xshift=0cm,yshift=0.4cm]
\tikzstyle{every node}=[scale=0.85]
\node(input_1)[input]{};
\node(Sptm)[encoder,right of=input_1,xshift=-1cm,yshift=1.6cm]{Speech pre-trained \\ model};
\node(Aa)[encoder,right of=Sptm,xshift=0cm,yshift=1.6cm]{Alignment \\ adapter};
\node(Sa)[encoder,right of=Aa,xshift=0cm,yshift=2.4cm]{Textual \\ adapter};
\node(encoder_t)[decoder,right of=Sptm,draw=gray,dashed,fill=gray!10,xshift=3.5cm,yshift=0.8cm]{ \textcolor{red}{Denoising}\\T-Encoder};
\node(decoder)[decoder,right of = encoder_t,xshift=3.4cm,yshift=0cm]{Decoder};
\node(input_2)[input,below of=decoder,xshift=-0.7cm, yshift=-2.4cm,draw=LightCoral,fill=LightCoral!10]{};

\node(input_s)[below of = Sptm,xshift=-0.5cm,yshift=-1.6cm]{\small{Speech ($s$)}};
\node(input_tx)[below of = encoder_t,xshift=-1.4cm,yshift=-2.4cm]{\small{Transcription ($t$)}};
\node(input_st)[below of = encoder_t,xshift=1.5cm,yshift=-2.4cm]{\small{Source text ($x$)}};
\node(input_tt)[below of = decoder,xshift=0.6cm,yshift=-2.44cm]{\small{Target text ($y$)}};

\draw[->,thick](Sptm.north)to(Aa.south);
\draw[->,thick](Aa.north)to(Sa.south);
\draw[->,thick]([yshift=-0.2cm]encoder_t.east)to([yshift=-0.2cm]decoder.west);
\draw[->,thick]([xshift=-1cm]input_1.north)to(Sptm.south);
\draw[->,thick]([xshift=1.9cm]input_1.north)to([xshift=-0.6cm]encoder_t.south);
\draw[->,thick]([xshift=-2.1cm]input_2.north)to([xshift=0.6cm]encoder_t.south);
\draw[->,thick]([yshift=-0.6cm]decoder.south)to(decoder.south);

\node(output_loss2)[above of = decoder,xshift=-0.5cm,yshift=2.3cm]{{$Loss_{\mathrm{MT}}$}};
\node(output_loss3)[above of = decoder,xshift=1.0cm,yshift=2.3cm]{{$Loss_{\mathrm{ST}}$}};
\node(output_loss4)[above of = Sa,xshift=2.5cm,yshift=0.5cm,color=red]{\small{$Loss_{\mathrm{contrastive}}$}};

\draw[->,thick]([xshift=-0.5cm]decoder.north)to(output_loss2.south);
\draw[->,thick]([xshift=1.0cm]decoder.north)to(output_loss3.south);
\draw[->,thick](Sa.north)to([yshift=0.4cm]Sa.north)
-|([xshift=-0.4cm,yshift=0.2cm]decoder.west)to([yshift=0.2cm]decoder.west);
\draw[-,thick,draw=red](Sa.east)-|(encoder_t.north);

\node(loss)[above of = Aa,xshift=1.4cm,yshift=1.3cm]{{$Loss_{\mathrm{ASR}}$}};
\draw[->,thick]([xshift=-0.54cm]loss.west)to(loss.west);
\draw[->,thick,draw=red]([yshift=-0.25cm]output_loss4.south)to([yshift=0.1cm]output_loss4.south);
\node(num_sptm2)[num,left of = Sptm,xshift=-1.4cm,yshift=0.6cm]{\small\bf{2}};
\node(num_sptm3)[num,left of = Sptm,xshift=-0.9cm,yshift=0.6cm]{\small\bf{3}};
\node(num_Aa2)[num,left of = Aa,xshift=-1.4cm,yshift=0.6cm]{\small\bf{2}};
\node(num_Aa3)[num,left of = Aa,xshift=-0.9cm,yshift=0.6cm]{\small\bf{3}};
\node(num_Sa2)[num,left of = Sa,xshift=-1.4cm,yshift=0.6cm]{\small\bf{2}};
\node(num_Sa3)[num,left of = Sa,xshift=-0.9cm,yshift=0.6cm]{\small\bf{3}};
\node(num_et1)[num,left of = encoder_t,xshift=-1.4cm,yshift=1.4cm]{\small\bf{1}};
\node(num_et2)[num,left of = encoder_t,xshift=-0.9cm,yshift=1.4cm]{\small\bf{2}};
\node(num_de1)[num,left of = decoder,xshift=-1.4cm,yshift=1.4cm]{\small\bf{1}};
\node(num_de3)[num,left of = decoder,xshift=-0.9cm,yshift=1.4cm]{\small\bf{3}};
\end{scope}
\begin{scope}[xshift=11.5cm,yshift=6.7cm,scale=0.5]
\tikzstyle{every node}=[scale=0.5]
\node(encoder_t)[decoder,minimum height=1.2cm,rounded corners=0.1cm,yshift=-.3cm,dashed,draw=gray,fill=gray!10]{ \textcolor{red}{Denoising}\\[-0.5ex] T-Encoder};
\node(decoder)[decoder,minimum height=1.2cm,rounded corners=0.1cm,right of = encoder_t,xshift=3.4cm,yshift=0cm]{Decoder};
\node(input)[input,below of=encoder_t,rounded corners=0.1cm,xshift=1.7cm, yshift=-1.3cm,draw=LightCoral,minimum width=6.2cm,fill=LightCoral!10]{};
\node(input_st)[below of = input,xshift=-1.4cm,yshift=0cm]{\small{Source text}};
\node(input_tt)[below of = input,xshift=1.6cm,yshift=0cm]{\small{Target text}};
\node(loss)[above of=decoder,yshift=1.2cm]{{$Loss_{\mathrm{MT}}$}};

\draw[->]([yshift=-0.45cm]encoder_t.south)to(encoder_t.south);
\draw[->]([yshift=-0.45cm]decoder.south)to(decoder.south);
\draw[->](encoder_t.east)to(decoder.west);
\draw[->](decoder.north)to(loss.south);
\node(backdrop)[rectangle,thick,dotted,rounded corners,minimum width=10cm,minimum height=3.9cm,xshift=0.5cm,yshift=-0.3cm,text centered,draw=blue!50]{};
\node(step1)[above of=backdrop,xshift=-3.5cm, yshift=0.1cm,text centered,align=left,font=\Large]{\LARGE{Step \quad \ :}\\[3ex] MT \\ Training};
\node(num)[num,right of = step1,xshift=0.55cm,yshift=1cm,scale=1.5]{\small\bf{1}};

\end{scope}
\begin{scope}[xshift=11.5cm,yshift=3.7cm,scale=0.5]
\tikzstyle{every node}=[scale=0.5]
\node(Sptm)[encoder,rounded corners=0.05cm,minimum height=0.5cm]{ Speech model};
\node(Aa)[encoder,above of=Sptm,rounded corners=0.05cm,yshift=0.8cm,minimum height=0.5cm]{\footnotesize{Alignment adapter}};
\node(Sa)[encoder,above of=Aa,rounded corners=0.05cm,yshift=0.8cm,minimum height=0.5cm]{Textual adapter};
\node(encoder_t)[decoder,dashed,rounded corners=0.1cm,right of = Sptm,minimum height=1.32cm,xshift=3.4cm,yshift=0.39cm,draw=gray,fill=gray!10]{\textcolor{red}{Denoising}\\[-0.5ex] T-Encoder};
\node(input)[input,below of=Sptm,rounded corners=0.1cm,xshift=1.7cm, yshift=-1cm,minimum width=6.2cm]{};
\node(input_st)[below of = input,xshift=-1.6cm,yshift=0cm]{\small{Speech}};
\node(input_tt)[below of = input,xshift=1.6cm,yshift=0cm]{\small{Transcription}};
\node(loss)[right of=Sa,xshift=3.2cm,yshift=0.4cm,color=red]{{$Loss_{\mathrm{contrastive}}$}};
\node(ASRloss)[right of=Sa,xshift=2.2cm,yshift=-0.25cm]{{$Loss_{\mathrm{ASR}}$}};
\draw[->]([xshift=-1.68cm,yshift=-0.15cm]ASRloss.west)to([xshift=-0.1cm,yshift=-0.15cm]ASRloss.west);
\draw[->]([yshift=-0.5cm]Sptm.south)to(Sptm.south);
\draw[->](Sptm.north)to(Aa.south);
\draw[->](Aa.north)to(Sa.south);
\draw[->]([yshift=-0.5cm]encoder_t.south)to(encoder_t.south);
\draw[-,color=red](Sa.east)-|([xshift=0.4cm]encoder_t.north);
\draw[->,color=red]([xshift=-0.5cm,yshift=-0.16cm]loss.south)to([xshift=-0.5cm,yshift=0.2cm]loss.south);
\node(backdrop)[rectangle,thick,dotted,rounded corners,minimum width=10cm,minimum height=3.9cm,xshift=0.5cm,yshift=0.3cm,text centered,draw=blue!50]{};
\node(step2)[above of=backdrop,xshift=-3.5cm, yshift=0.1cm,text centered,align=left,font=\Large]{\LARGE{Step \quad \ :}\\[2ex] ASR \\ Training};
\node(num)[num,right of = step2,xshift=0.55cm,yshift=0.9cm,scale=1.5]{\small\bf{2}};
\end{scope}
\begin{scope}[xshift=11.5cm,yshift=1.1cm,scale=0.5]
\tikzstyle{every node}=[scale=0.5]
\node(Sptm)[encoder,rounded corners=0.05cm,minimum height=0.5cm]{ Speech model};
\node(Aa)[encoder,above of=Sptm,rounded corners=0.05cm,yshift=0.8cm,minimum height=0.5cm]{\footnotesize{Alignment adapter}};
\node(Sa)[encoder,above of=Aa,rounded corners=0.05cm,yshift=0.8cm,minimum height=0.5cm]{Textual adapter};
\node(decoder)[decoder,decoder,rounded corners=0.1cm,right of = Sptm,minimum height=1.32cm,xshift=3.4cm,yshift=0.39cm]{Decoder};
\node(input)[input,below of=Sptm,rounded corners=0.1cm,xshift=1.7cm, yshift=-1cm,minimum width=6.2cm,draw=color1,fill=color1!10]{};
\node(input_st)[below of = input,xshift=-1.6cm,yshift=0cm]{\small{Speech}};
\node(input_tt)[below of = input,xshift=1.6cm,yshift=0cm]{\small{Target text}};

\node(loss)[above of=decoder,xshift=0cm,yshift=1.3cm]{{$Loss_{\mathrm{ST}}$}};
\draw[->]([yshift=-0.5cm]Sptm.south)to(Sptm.south);
\draw[->]([yshift=-0.5cm]decoder.south)to(decoder.south);
\draw[->](Sptm.north)to(Aa.south);
\draw[->](Aa.north)to(Sa.south);
\draw[->](Sa.east)to([xshift=0.4cm]Sa.east)|-(decoder.west);
\draw[->](decoder.north)to(loss.south);
\node(backdrop)[rectangle,thick,dotted,rounded corners,minimum width=10cm,minimum height=3.9cm,xshift=0.5cm,yshift=0.3cm,text centered,draw=blue!50]{};
\node(step3)[above of=backdrop,xshift=-3.5cm, yshift=0.1cm,text centered,align=left,font=\Large]{\LARGE{Step \quad \ :}\\[2ex] ST \\ Training};
\node(num)[num,right of = step3,xshift=0.55cm,yshift=0.9cm,scale=1.5]{\small\bf{3}};
\end{scope}
\end{tikzpicture}

\caption{The end-to-end speech translation architecture with a text encoder. Circled numbers indicate training steps. The denoising text-encoder (T-encoder) will be dropped during Step 3.}
\label{fig:model architecture}

\end{figure*}
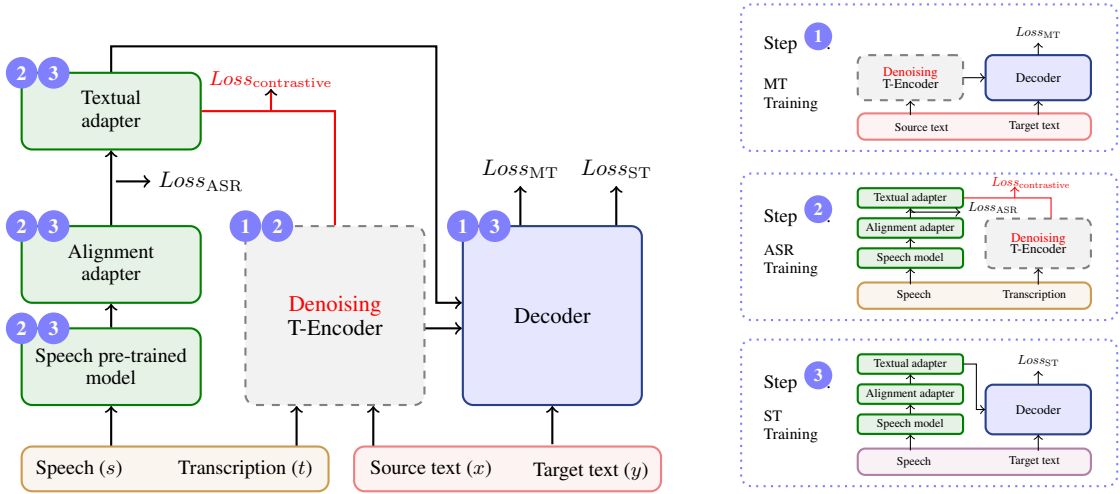

Here, we take a further step towards more effective use of both labeled and unlabeled data in ST. We claim that the source-language text encoder plays an important role in leveraging ASR and MT although it is not involved in standard end-to-end ST. We then develop a method (named as Multi-Step Pre-training for Speech Translation, or MSP-ST for short) to expose the text encoder to both ASR and MT learning processes, and force them to assist each other. Having the text encoder as the bridge between ASR and MT is perhaps helpful: the result ST system can learn acoustic and textual encoding simultaneously (see Figure \ref{fig:model-types}). Note that such a design also addresses the role mismatch problem wherein the pre-trained ASR encoder does not behave like what the target-language decoder expects \cite{wang2020bridging,xu-etal-2021-stacked}. To our knowledge, this is the first to discuss the problem in large-scale pre-training with all ASR, MT and ST data on end-to-end ST tasks.

Another improvement is that we denoise the text encoder so that it is robust to the noisy transcription-like input. In this way, the text encoder can deal with both the normal text and the transcription. This is beneficial when the text encoder is used to supervise the learning of the ST encoder, where the speech-to-transcription data is the input.

We implement our method in a Transformer-based ST system. On the MuST-C and LibriSpeech tasks, it outperforms very strong baselines significantly. It achieves BLEU scores of 30.0 and 40.6 on the MuST-C En-De and En-Fr data and a BLEU score of 21.4 on the LibriSpeech En-Fr data. These results are new state-of-the-arts on these tasks. The performance is even comparable with that of the unrestricted MT system on the LibriSpeech task. We implement our method in the submitted system for IWSLT 2022 offline speech translation task \cite{zhang-etal-2022-niutranss}.

\section{Method}

Our ST model is a standard encoder-decoder model, following the Transformer model \cite{vaswani2017attention}. The encoder reads a sequence of source-language acoustic signals, and the decoder produces a sequence of target-language words. Broadly speaking, like any encoder-decoder model, one can train this architecture in a standard pre-training + fine-tuning fashion \cite{lewis-etal-2020-bart}. For example, the encoder is pre-trained by pure acoustic data \cite{baevski2020wav2vec}, and/or enhanced by training an ASR encoder on speech-to-transcription data. Likewise, the decoder is initialized by pre-trained models (for either the word embedding component or the whole decoding network). The final ST model is tuned on the labeled data, i.e., pairs of speech and translation.

But such a model does not accept source-language text as input, and it is non-trivial to learn the model on source-language text data. One way to use textual input is to have a sub-model, implicit or explicit, to introduce source-language text signals into the ST model. To this end, we develop a text encoder on the source-language side in addition to the ST encoder. In pre-training, it works with both the ST encoder and decoder. After that, the text encoder is absent, and the ST model is tuned and then used for inference, as usual.

\begin{table}[t]
\centering
\setlength{\tabcolsep}{4pt}
\begin{tabular}{r|l|l|l|l}
Order & Name & Data & Trained & Training \\
          &            & Type & Model   &  \\ \hline
  & & $s$ & s-enc. & pre-train \\
1 & Init. & $x$ & t-enc. & pre-train \\
  & & $y$ & dec. & pre-train \\ \hline
2 & MT & $(x,y)$ & t-enc + dec. & pre-train \\ \hline
3 & ASR & $(s,t)$ & s-enc + t-enc. & pre-train \\ \hline
4 & ST & $(s,y)$ & s-enc. + dec. & fine-tune
\end{tabular}

\caption{Data types used in training. s-enc. = ST encoder, t-enc = text encoder, and dec. = ST decoder.}
\label{tab:data-type}

\end{table}

Formally, let $s$ be an acoustic signal sequence, $t$ be a transcription of $s$, $x$ be a source-language word sequence, and $y$ be a target-language word sequence. There are many choices to build different types of training data. For example, $(s, y)$ is the standard ST data, $(x,y)$ is the MT data, $x$ is the monolingual data. Table \ref{tab:data-type} shows the data types used here, ordered by the training pipeline of our method. Note that the term ``pre-training'' is used in many different ways. In this paper, the term refers to any training process other than the final tuning of the ST model on $(s, y)$\footnote{Training on $(x,y)$ and $(s,t)$ is actually a ``tuning'' process on the initialized/pre-trained model. Here we call them pre-training to avoid the misuse of ``tuning'' because it is typically used when tuning the model on the labeled target-task data.}.

Another note on notation. Not all these sequences are required to come in pairs. For example, $x$ in the monolingual data might not appear in the MT data. Here we use this notation to emphasize what type of data is used in training, but not the actual data.

At the heart of our system is a design to guide the ST model via textual information. Two intuitions form the basis of this work:
\begin{itemize}
\item The text encoder can supervise the training of the ST encoder so that the behavior of the ST encoder is more consistent with that of a standard MT encoder.
\item The text encoder can be robust to ASR noise, and can accept transcription as input.
\end{itemize}
To make use of these intuitions, we improve the ST encoder and develop a contrastive training method to incorporate the text encoder into the ASR-based training. Beyond this, we propose a denoising method to learn a text encoder that is robust to either normal text or transcription.
\subsection{ASR Training with the Text Encoder}

ST encoders in general share a similar model structure with ASR encoders. An advantage of the ASR-based design for ST encoders is that it is better suited for processing acoustic signals and off-the-shelf pre-trained acoustic models are straightforwardly available to ST. However, the ASR-like encoder does not work with the target-language text decoder because the decoder wants text-friendly encoding instead of the acoustic encoding \cite{dong2021listen,xu-etal-2021-stacked}. A way to address this modality-inconsistency issue is to stack adapters on top of the acoustic model. Thus, the system can learn to transform from one modality to another. However, it remains undesirable that the supervision of the encoder is only from the decoder and the vast number of source-language sentences are ignored.

We propose to use the text encoder to supervise the training of the ST encoder. See Figure \ref{fig:model architecture} for the model architecture. The core design is the adapters for the ST encoder and the contrastive learning for the two encoders.

\subsubsection{Adapters for ST Encoding}

For ST encoding, Connectionist Temporal Classification (CTC) based training is necessary for state-of-the-art performance \cite{graves2006connectionist}. A common way is to add the CTC-based loss to the acoustic model. Then, an optional adapter can be used to map the acoustic model output to representations that the text decoder prefers \cite{xu-etal-2021-stacked}, or shrinks the length of the acoustic model output \cite{li-etal-2021-multilingual}.

In our preliminary experiments, we found that it was not easy to do alignment in CTC-based training due to the big length difference between the acoustic model output and the word sequence. Thus, we propose an alignment adapter and place it between the acoustic model and the CTC-based loss. The adapter consists of $n$ convolutional networks to shorten the sequence and a Conformer layer \cite{9413423} to filter the down-sampling output. To make a stronger correlation with the text encoder, we share the same vocabulary and the output layer to predict each word in the representation space of the textual model when generating the CTC path. This way forces the acoustic representation space to align to that of the text encoder.

Another encoding network (call it textual adapter) is stacked upon the alignment adapter. It consists of a single self-attention layer. We add the position embedding before feeding the feature into this adapter to fuse location information. The textual adapter is intended to reduce the impact of blank noise and produces a more text encoder-like output, which is better suited for the input of the decoder.

\begin{table*}[t]
    \centering
    \small
    \setlength{\tabcolsep}{3.0mm}{
    \begin{tabular}{lccccccc}
    \toprule
     Models & Speech&Text & ASR & MT & \makecell[c]{MuST-C\\En-De} & \makecell[c]{MuST-C\\En-Fr} & \makecell[c]{LibriSpeech\\En-Fr}\\
    \midrule
    Unrestricted MT \cite{xu-etal-2021-stacked} &-&-&-&-& 31.1&$\textrm{41.9}^{*}$&21.3 \\
    \hline
    Transformer \cite{wang-etal-2020-fairseq}&-&-&-&-&22.7&32.9&16.7\\
    FAT-ST (Big) \cite{pmlr-v139-zheng21a}&$\checkmark$&$\checkmark$&$\checkmark$&$\checkmark$&25.5&-&- \\
    VggTLarge \cite{pino20_interspeech}&$\checkmark$&-&$\checkmark$&-&25.6&-&- \\
    LUT \cite{dong2021listen} &-&$\checkmark$&$\checkmark$&-&-&-&18.3\\
    Chimera \cite{han-etal-2021-learning}&$\checkmark$&-&-&$\checkmark$& 26.3&35.6&19.4 \\
    JT \cite{tang-etal-2021-improving} &-&-&-&$\checkmark$&26.8&37.4&-\\
    LNA-ED-Adapt \cite{gallego2021end} &$\checkmark$&$\checkmark$&$\checkmark$&$\checkmark$&27.3&-&-\\
    XSTNET \cite{ye2021end}&$\checkmark$&-&-&$\checkmark$&27.8 &38.0&-\\
    SATE \cite{xu-etal-2021-stacked}&-&-&$\checkmark$&$\checkmark$&28.1&-&20.8\\
    TCN \cite{9414703}&-&-&$\checkmark$&$\checkmark$& 28.9 &-&-\\
    STPT \cite{tang2022unified}&$\checkmark$&-&$\checkmark$&$\checkmark$ &-&39.7&-\\
    \hline
    Baseline &$\checkmark$&$\checkmark$&$\checkmark$&$\checkmark$& 27.5&38.6 &20.8\\
    MSP-ST &$\checkmark$&$\checkmark$&$\checkmark$&$\checkmark$ & 30.0 &40.6 &\textbf{21.4} \\
    MSP-ST-H$^{\dag}$ &$\checkmark$&$\checkmark$&$\checkmark$&$\checkmark$ & \textbf{30.2} &\textbf{40.8} &- \\
    \bottomrule
    \end{tabular}}
    \caption{Performance on different data set. The baseline is LNA \cite{li-etal-2021-multilingual} and add an additional adapter \cite{gallego2021end}. $^{*}$ represents that we reproduce the result. $^{\dag}$ uses HuBERT \cite{9585401} as the pre-trained acoustic encoder.}
    \label{main_task}
\end{table*}

\subsubsection{Contrastive Training}
\label{sec:contrastive-training}

We train the ST encoder with the text encoder in addition to the supervision signal from the decoder side. This is a step before we fine-tune the ST model. Here we choose contrastive training as a way to connect the ST encoder and the text encoder. More formally, let $\mathcal{A}(s)$ be the output of the ST encoder given the speech $s$, and $\mathcal{M}(t)$ be the output of the pre-trained text encoder given the transcription $t$. Given a set of training samples $\{(s_i,t_i)\}$, the loss function of the contrastive training is defined to be:
\begin{equation}
    \small
    \mathcal{L}_{\mathrm{CL}}=-\sum_{(s_i,t_i)} \log \frac{e^{\pi\left(\mathcal{A}\left(s_i\right), \mathcal{M}\left(t_i\right)\right)/\tau }} {\sum_{t_j:j \neq i} e^{\pi\left(\mathcal{A}\left(s_i\right), \mathcal{M}\left(t_j\right)\right)/\tau}} \label{eq:contrastive-learning}
\end{equation}

\noindent where $\pi(\cdot,\cdot)$ is a function that computes the similarity of the input vectors. Here we choose the cosine function for $\pi(\cdot,\cdot)$ and average pooling the two sequence representations. $\tau$ is a scaler to control the sharpness of the function output. For each $s_i$, we have its labeled transcription to form a positive sample $(s_i, t_i)$. Also, we use transcriptions other than $t_i$ (i.e., $t_j$ for $j \neq i$) to form negative samples. Eq. \ref{eq:contrastive-learning} distinguishes the positive sample from the negative samples (i.e., $\{(s_i,t_j)|j \neq i\}$).
Thus, $\mathcal{A}(s_i)$ would be close to $\mathcal{M}(t_i)$ and far way from other $\mathcal{M}(t_j)$.

For more diverse training samples, we decode a transcription $t'_i$ by keeping blank labels in the output of the alignment adapter. For $(s_i,t'_i)$, we compute a loss $\mathcal{L'}_{\mathrm{CL}}$ as in Eq. \ref{eq:contrastive-learning}. The final loss function of ASR training is defined as:
\begin{equation}
\small
\mathcal{L}_{\mathrm{ASR}} =  \mathcal{L}_{\mathrm{CTC}} + \alpha ( \beta  \mathcal{L}_{\mathrm{CL}} + (1- \beta) \mathcal{L}_{\mathrm{CL}}^{'}) \label{eq:asr-loss}
\end{equation}

\noindent where $\mathcal{L}_{\mathrm{CTC}}$ is the CTC loss \cite{watanabe2017hybrid} which is widely used in ST task \cite{wang2020bridging,dong2021listen,xu-etal-2021-stacked}, and $\alpha$ and $\beta$ are coefficients for interpolation.

For guiding the textual adapter by contrastive training, the text encoder needs to suit the decoder and be well-trained. So we run the ASR training process after the MT training process and freeze the text encoder during ASR training.

\subsection{Denoising Text Encoder}

There are two jobs for the text encoder: 1) encoding real source-language sentences in MT training, 2) encoding transcriptions in ASR training.


As MT training is prior to ASR training, the text encoder is primarily trained to address the first point. This is potentially undesirable for a reason: in ASR training, the input of the text encoder is a transcription, which often contains symbols that never appear in MT data. The input will be more noisy if we use self-generated transcriptions in training (see Eq. \ref{eq:asr-loss}).

We use denoising methods for a robust text encoder, of which the simplest one is to use a denoising autoencoder (DAE) to take noise into account \cite{lewis-etal-2020-bart}. Here we choose mBART as the initial model \cite{liu-etal-2020-multilingual-denoising} for its potential cross-lingual ability. It is complicated to update mBART for introducing ASR-related noise (such as blank symbols) into DAE training. We therefore further denoise the encoder in the MT training phase to make a Silence Insensitive DAE (SIDAE). Our method is inspired by Consistency Regularization \cite{Zhang2020Consistency}. In consistency regularization, a ``good'' model should be less sensitive to perturbation on the input. We design a perturbation function $g(\cdot)$ that randomly adds blank symbols into source-language sentences. The size of adding blank is decided by the coefficient $r$ multiply the length of sentence. For each sentence pair $(x,y)$, we expect that the MT system can produce a correct prediction given both $x$ and $g(x)$ as input. The loss function on a set of source-text and target-text pairs $\{x_i,y_i\}$ is given by:


\begin{equation}
\small
\mathcal{L}_{\mathrm{MT}} = -\sum_{(x_i,y_i)} \log \textrm{P}(y_i \mid x_i) + \log \textrm{P}(y_i \mid g(x_i))
\end{equation}

\noindent where $\textrm{P}(y \mid \cdot)$ is the MT system consisting of the text encoder and the text decoder.

\section{Experiments}
\subsection{Data}
We run our experiments on English to German (En-De) and English to French (En-Fr) translation tasks. For speech data, we use the Librilight \cite{kahn2020libri} which consists of about 60k hours of unlabelled speech. For text data, we follow \citet{liu-etal-2020-multilingual-denoising}'s work which covers 25 languages.

We use LibriSpeech 960 hours \cite{7178964} to train the pre-trained acoustic model on the English ASR task. To adapt the DAE model to the MT task, we use the Opensubtitle En-De and WMT14 En-Fr datasets respectively. The final data consists of 18M sentence pairs for the En-De translation. For En-Fr translation, we extract 10M sentence pairs from the WMT14 En-Fr data, following \citet{xu-etal-2021-stacked}'s work. We use sentencepiece to segment the untokenized text into sub-words\footnote{https://github.com/google/sentencepiece}. The sentencepiece model and the vocabulary are the same as in \citet{liu-etal-2020-multilingual-denoising}'s work and we remove words which do not appear in all the corpora. The vocabulary size is set to 32K for the MuST-C tasks \cite{di-gangi-etal-2019-must} and 25K for the LibriSpeech En-Fr task 
\cite{kocabiyikoglu-etal-2018-augmenting}.

The MuST-C En-De and En-Fr tasks provide 400 hours and 484 hours speech data respectively. For the LibriSpeech En-Fr task, the size of the training set is 100 hours. Readers can refer to Appendix for more details about the information and processing of data.

\subsection{Model settings}
We implement our systems by the Fairseq toolkit \cite{ott2019fairseq}. For pre-training of unlabeled speech data, we use the open-source wav2vec2 model. For the DAE model, we also utilize the open-source mBART.CC25 model.
For a stronger baseline, we re-implement the LNA method \cite{li-etal-2021-multilingual}. Following \citet{gallego2021end}'s work, we add an adapter \cite{bapna-firat-2019-simple} to  mitigate the gap between the acoustic and textual model. We use speech as input for our pre-trained model.

For pre-training of SIDAE, we set the coefficient $r$ to 0.3.
We stop training until the perplexity converges on the validation set. For the alignment adapter, the size of the convolutional layer $n$ is set to 3. For each Conformer layer, there are 1,024 hidden states, 16 attention heads and 4,096 FFN hidden states. We freeze the pre-trained acoustic model in the first 5,000 training steps to warm up the two adapters. The $\tau$ and $\alpha$ are set to 0.1 and 0.3. The initial value of $\beta$ is 1. It then decreases by 0.1 per 5,000 steps until 0.
For fine-tuning on the ST task, we use the Adam optimizer with $\beta_{1}=0.9$ and $\beta_{2}=0.98$. We use dropout ($p=0.1$) and label smoothing ($p=0.1$) for robust training. We early stop the training if the last five checkpoints do not improve. For training settings of unrestricted MT, we follow \citet{xu-etal-2021-stacked}'s work.

When evaluating the model, we average the weight of the last five checkpoints. For inference, the beam size is set to 4 and the length penalty is set to 1.0. We use SacreBLEU \cite{post-2018-call} to evaluate the performance. Following previous work, we report the case-sensitive score for the MuST-C En-De and En-Fr tasks and case-insensitive score for the LibriSpeech En-Fr task.

\begin{table}[t]
    \centering
    \small
    \setlength{\tabcolsep}{4.0mm}{
    \begin{tabular}{lccc}
    \toprule
    Model & En-De & En-Fr\\
    \midrule
    Baseline & 27.5&38.6 \\
    \ \ + Alignment adapter &28.1&38.8  \\
    \ \ + Textual adapter, CL&29.1&39.8  \\
    \ \ + KDCL &29.5&40.2 \\
    \ \ + SIDAE &30.0&40.6 \\
    \bottomrule
    \end{tabular}
    }
    \caption{Ablation study on the MuST-C tasks.}
    \label{Ablaiton}
\end{table}

\subsection{Results}
Table \ref{main_task} shows our experimental results. We see, first of all, that our baselines which utilize all types of data are very strong and achieve the SOTA performance on LibriSpeech En-Fr tasks. While on the En-De task, the baseline fails to outperform the methods without using unlabeled data \cite{xu-etal-2021-stacked}. This proves previous result that using unlabeled data has more room for improvement.

Our method gains remarkable improvements on two MuST-C tasks compared with the baseline and achieves the SOTA results without using any ST data-augmentation method. Though our method only gains a +0.6 BLEU improvement on the LibriSpeech En-Fr task, it is comparable with the MT baseline.
Compared with \citet{xu-etal-2021-stacked}'s work, our method shows a +1.9 higher BLEU score by using additional unlabeled data. In particular, we use much less labeled data compared with TCN \cite{indurthi2020end} and still yields a 1.1 BLEU improvement. This also verifies the potential of unlabeled data in ST. Compared with the STPT \cite{tang2022unified}, we utilize the additional monolingual text data to build the denoising model and less MT data. Our method gains +0.9 BLEU improvement and shows the importance of denoising ability to the ST model. The MSP-ST achieves a 2.5 BLEU improvement compared with the baseline \cite{li-etal-2021-multilingual} due to the efficient use of unlabeled speech and text data.

\section{Analysis}
\subsection{Ablation Study}
We replace the adapters in the baseline system with our alignment adapter. Table \ref{Ablaiton} shows that the alignment adapter can achieve better performance. It indicates our alignment adapter is a more effective way to convert the representation space of the acoustic model to text model. Then, we introduce our textual adapter into the system and align it with the output of the DAE encoder by the contrastive loss. The results show that the textual adapter is the important for satisfactory performance. Also, this result confirms that the semantic conversion and denoising methods are important for ST. We introduce $L_\textrm{CL}^{'}$ (denoted as KDCL) into training (see Section \ref{sec:contrastive-training}). It shows that knowledge distillation can reduce the difficulty of semantic learning. The advances brought by the textual adapter and KDCL are the same apparently on the two tasks because the methods improve the acoustic model and use the similar speech data.
We finally use SIDAE to replace DAE and mitigate the impact of blank noise for the textual adapter. As expected, it helps. The final results achieve new SOTA performance on the two MuST-C tasks.

\begin{table}[t]
    \centering
    \small
    \setlength{\tabcolsep}{3.0mm}{
    \begin{tabular}{lrrc}
    \toprule
    Model & ST data & Utterances &Test \\
    \midrule
    Transformer & 65h & 39K & 6.4 \\
    Transformer &  400h & 230K&  22.7 \\
    MSP-ST & 10h & 5K & 15.9 \\
    MSP-ST & 65h & 39K & 24.3 \\
    MSP-ST & 400h & 230K& 30.0 \\
    \bottomrule
    \end{tabular}
    }
    \caption{Sample efficiency on the MuST-C En-De task.}
    \label{fewshot}
\end{table}

\subsection{Effect of Denoising}
\begin{figure}[t]
  \setlength{\abovecaptionskip}{0.1cm}
  \centering
  \begin{tikzpicture}
  \centering
    \scriptsize{
    \begin{axis}[
      at={(0,0)},
      ymajorgrids,
      grid style=dashed,
      legend style={at={(0.02,0.65)}, anchor=south west},
      legend cell align={left},
      ybar=3pt,
      enlarge x limits=0.5,
      xtick align=inside,
      height=.24\textwidth,
      width=.27\textwidth,
      bar width=1.0em,
      xlabel={(a) \ Performance on clean and\\noisy test set},
      ylabel={BLEU},
      symbolic x coords={{1}, {2}},
      xtick=data,
      nodes near coords align={vertical},
      ymin=24,
      ymax=32,
      ytick={25,28,31},
      xticklabels={Clean test,Noisy test},
      legend entries={DAE,SIDAE},
      ylabel style={yshift=-3em},xlabel style={yshift=0.3em,align=center},
      yticklabel style={/pgf/number format/fixed,/pgf/number format/fixed zerofill,/pgf/number format/precision=1,rotate=90},
      ]
          \addplot[fill=red!30, draw=red, area legend] coordinates {({1},28.5) ({2},25.0) };

          \addplot[fill=blue!30, draw=blue, area legend] coordinates {({1},28.8) ({2},28.9) };

    \end{axis}
    }

    \scriptsize{
    \begin{axis}[
      at={(15.6em,0)},
      ymajorgrids,
      grid style=dashed,
      legend style={at={(0.41,0.54)}, anchor=south west},
      legend cell align={left},
      ybar,
      enlarge x limits=0.5,
      xtick align=inside,
      height=.24\textwidth,
      width=.27\textwidth,
      bar width=0.8em,
      xlabel={(b) \ Performance on different\\ratio of noise},
      ylabel={BLEU},
      symbolic x coords={{1}, {2}},
      xtick=data,
      nodes near coords align={vertical},
      ymin=27,
      ymax=31.5,
      ytick={27.5,29.2,31},
      xticklabels={Low noise,High noise},
      legend entries={Baseline,w/o CL,CL},
      ylabel style={yshift=-3em},xlabel style={yshift=0.3em,align=center},
      yticklabel style={/pgf/number format/fixed,/pgf/number format/fixed zerofill,/pgf/number format/precision=1,rotate=90},
      ]
          \addplot[fill=red!30, draw=red,area legend] coordinates {({1},27.4) ({2},27.5) };

          \addplot[fill=blue!30, draw=blue,area legend] coordinates {({1},28.5) ({2},27.9) };

          \addplot[fill=teal!30, draw=teal,area legend] coordinates {({1},30.2) ({2},29.1) };

    \end{axis}
    }

\end{tikzpicture}
\caption{Comparison of denoising or not on MuST-C En-De test.}\label{denosieperformance}
\end{figure}
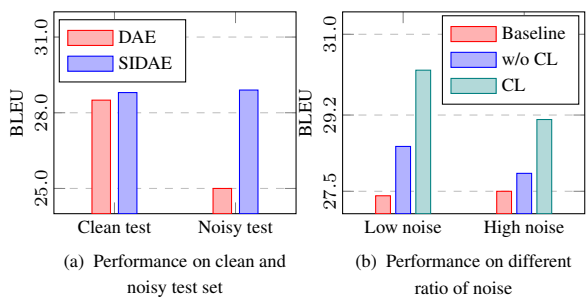

Figure \ref{denosieperformance} (a) compares BLUE scores of DAE and SIDAE. We use the blank label to replace punctuations in the source text to get the noisy test set. The performance on the clean test is almost the same. The modest improvement of the SIDAE model may be due to the stronger generalization ability by noisy training. When the test text contains many blank labels, the vanilla DAE model performs worse while the SIDAE is robust to the noise. To explore the denosing influence on ST, we split the test set into two sets according to whether the blank label ratio is higher than 0.3. Figure \ref{denosieperformance} (b) shows the performance of different systems on the test sets. Here ``w/o CL'' and ``CL'' mean that the textual adapter is trained without and with contrastive loss. Its improvement is modest on the high noise test set, while our textual adapter achieves a bigger BLEU improvement.

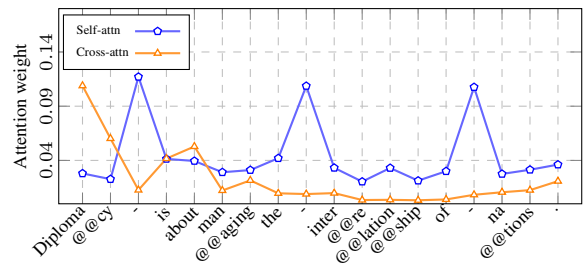
\begin{figure}[t]
  \setlength{\abovecaptionskip}{0.1cm}
  \centering
  \begin{tikzpicture}
  \scriptsize{
    \begin{axis}
    [
	  at={(0,0)},
      ymajorgrids,
      xmajorgrids,
      grid style=dashed,
      width=.48\textwidth,
      height=.235\textwidth,
      enlarge x limits=0.05,
      legend style={at={(0.1,0.55)}, anchor=south west},
      legend cell align={left},
      ylabel={\scriptsize{Attention weight}},
      ylabel style={yshift=-3em},xlabel style={yshift=0.0em},
      yticklabel style={/pgf/number format/fixed,/pgf/number format/precision=2,/pgf/number format/fixed zerofill,rotate=90},
      ymin=0.,ymax=0.18,
      scaled y ticks=false,
      ytick={0.04,0.09,0.14},
      xtick={1,2,3,4,5,6,7,8,9,10,11,12,13,14,15,16,17,18},
      xticklabels={Diploma,@@cy,-,is,about,man,@@aging,the,-,inter,@@re,@@lation,@@ship,of,-,na,@@tions,.},
      x tick label style={
            rotate=45,
            anchor=east,
            xshift=0.05cm,
            yshift=-0.05cm
        },
      legend style={yshift=10pt,xshift=-2.6em, legend plot pos=right,font={\tiny},cells={anchor= west}}
      ]
      \addplot[blue!60,mark=pentagon*,mark size=1.5pt,thick,mark options={fill=white,draw=blue,line width=0.5pt}] coordinates {(1,0.0279826)
      (2,0.02270425)
      (3,0.11705522)
      (4,0.0413757)
      (5,0.03946775)
      (6,0.02906379)
      (7,0.03108248)
      (8,0.04201537)
      (9,0.10864827)
      (10,0.03316183)
      (11,0.02042497)
      (12,0.03299978)
      (13,0.02132071)
      (14,0.03002064)
      (15,0.10758135)
      (16,0.027514)
      (17,0.03137828)
      (18,0.0361344)
      };
      \addlegendentry{\scalebox{.8}{Self-attn}}

      \addplot[orange!80,mark=triangle*,,mark size=1.5pt,thick,mark options={fill=white,draw=orange,line width=0.5pt}] coordinates {
      (1,0.10885818)
      (2,0.06019562)
      (3,0.01268586)
      (4,0.04167344)
      (5,0.05282197)
      (6,0.01225204)
      (7,0.02158166)
      (8,0.00965157)
      (9,0.0089652 )
      (10,0.00983812)
      (11,0.00345364)
      (12,0.00364504)
      (13,0.00322703)
      (14,0.003992)
      (15,0.00829033)
      (16,0.01062945)
      (17,0.01257407)
      (18,0.02085776)
      };
      \addlegendentry{\scalebox{.8}{Cross-attn}}

    \end{axis}}
    \end{tikzpicture}
    \caption{Blank self-attention and other word cross attention. ``-'' represents the blank token.}\label{Denoising}
\end{figure}

We further explore why the SIDAE model is not impacted so much by blank symbols. As Figure \ref{Denoising} shows, the self-attention weight of blank label focus on all blank labels, which means that the output of this position is only with a blank message and it is easy to be recognized in the cross-attention module. The attention weights of cross-attention confirm our conjecture, the position of silent speech has a very low weight. Thus, the blank noise can not affect the interference process. In the rest of this paper, we use the SIDAE model to guide the textual adapter.


\subsection{Impact of Alignment Adapter}
 Here we show the effectiveness of the alignment adapter. We select several high-frequency words from a random sentence and calculate the cosine similarities of words between the acoustic model and textual model in Figure \ref{similarity_alignment}. For cross-modal (cross-lingual) similarity, it’s the similarity between the output of the alignment adapter and source (target) token embedding. The baseline model does not consider the alignment of the acoustic model and the text encoder. Both the cross-modal and cross-lingual similarities are almost around zero. The inconsistency of representation space aggravates the gap between acoustic and textual models. The alignment adapter boosts the alignment between the two models and can reduce the difficulty of contrastive learning because the adapter does not need to consider the transfer of the representation space. Because of the cross-lingual nature of multilingual DAE, the cross-language alignment can also better facilitate language transfer.

\subsection{Impact of Textual Adapter}
\begin{figure}[t]
  \setlength{\abovecaptionskip}{0.12cm}
  \centering
  \begin{tikzpicture}
  \scriptsize{
    \begin{axis}
    [
	 at={(0,0)},
      ymajorgrids,
      xmajorgrids,
      grid style=dashed,
      width=.48\textwidth,
      height=.23\textwidth,
      enlarge x limits=0.05,
      legend style={at={(0.76,0.52)}, anchor=south west},
      ylabel={\scriptsize{Attention weight}},
      ylabel style={yshift=-3em},xlabel style={yshift=0.5em},
      yticklabel style={/pgf/number format/precision=2,/pgf/number format/fixed zerofill,rotate=90},
      ymin=-0.,ymax=0.4,
      ytick={0.1,0.2,0.3},
      xtick={1,2,3,4,5,6,7,8,9,10,11,12,13,14,15,16,17,18,19,20,21,22},
      xticklabels={im,going,going,-,to,talk,-,today,-,about,about,-,-,energy,energy,-,-,and,-,climate,-,-},
      x tick label style={
            rotate=45,
            anchor=east,
            xshift=-0.1cm,
            yshift=-0.15cm
        },
      legend style={yshift=10pt,xshift=-2.7em, legend plot pos=right,font={\tiny},cells={anchor= west}}
      ]
      \addplot[blue!60,mark=pentagon*,mark size=1.5pt,thick,mark options={fill=white,draw=blue,line width=0.5pt}] coordinates {
      (1, 0.02738269)
      (2, 0.04657348)
      (3, 0.09601289)
      (4, 0.03567865)
      (5, 0.04541012)
      (6, 0.02829216)
      (7, 0.02013602)
      (8, 0.00567486)
      (9, 0.0078078)
      (10,0.24318647)
      (11,0.16863865)
      (12,0.00512379)
      (13,0.0019645)
      (14,0.00351414)
      (15,0.00333768)
      (16,0.00342422)
      (17,0.00302563)
      (18,0.00160283)
      (19,0.00384765)
      (20,0.01975905)
      (21,0.03610331)
      (22,0.19350344)
      };

      \addlegendentry{\scalebox{.8}{Alignment adapter}}

      \addplot[orange!80,mark=triangle*,,mark size=1.5pt,thick,mark options={fill=white,draw=orange,line width=0.5pt}] coordinates {
      (1, 0.31183761)
      (2, 0.02202013)
      (3, 0.06250495)
      (4, 0)
      (5,0)
      (6, 0.06249999)
      (7, 0.00808627)
      (8, 0.05364618)
      (9, 0.00000127)
      (10, 0.16074619)
      (11,0.15903383)
      (12,0.00076234)
      (13,0)
      (14,0)
      (15,0)
      (16,0)
      (17.0)
      (18,0)
      (19,0)
      (20,0)
      (21,0)
      (22,0.15886118)
      };

      \addlegendentry{\scalebox{.8}{Textual adapter}}

    \end{axis}}
    \end{tikzpicture}
    \caption{The self-attention weights of alignment adapter and textual adapter.}\label{CL}
\end{figure}
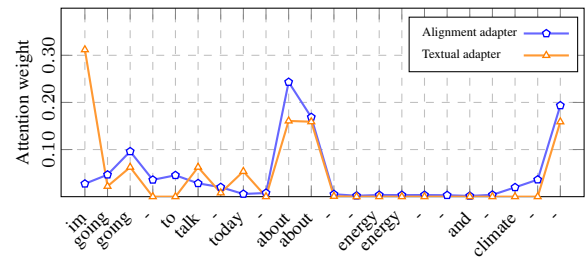

To study the impact of the textual adapter, we compare the attention wights between the alignment adapter and the textual adapter. Figure \ref{CL} shows that the textual adapter is helpful in adapting the ST encoder to a text-friendly encoder. The weight in the 1st position shows that the textual adapter learns information which may be unimportant for the cross-modal stage. This proves the difference between the acoustic model and textual model in encoding the input. Further, the adapter focuses more on the first position which is more important at the stage of translation. This indicates that the adapter learns something better suited to the MT model. Figure \ref{CL} also shows that in many blank positions, the weights are lower than those of the alignment adapter.

\begin{figure}[t]
  \setlength{\abovecaptionskip}{0.1cm}
  \centering
  \begin{tikzpicture}
  \centering
    \scriptsize{
    \begin{axis}[
      name=border,
      at={(0,0)},
      ymajorgrids,
      grid style=dashed,
      legend style={at={(0.3,1.05)}, anchor=south west},
      legend columns=-1,
      ybar=5pt,
      enlarge x limits=0,
      xtick align=inside,
      height=.24\textwidth,
      width=.19\textwidth,
      bar width=0.8em,
      xlabel={(a) \ Similarity between the\\acoustic and textual encoder},
      ylabel={Similarity},
      symbolic x coords={{1}},
      xtick=data,
      nodes near coords align={vertical},
      ymin=0,
      ymax=0.5,
      ytick={0.10,0.25,0.4},
      xticklabels={},
      legend entries={Baseline,w/o CL,Textual adapter,MSP-ST},
      ylabel style={yshift=-3em},xlabel style={yshift=1.8em,align=center},
      yticklabel style={/pgf/number format/fixed,/pgf/number format/fixed zerofill,/pgf/number format/precision=2,rotate=90},
      ]
          \addplot[fill=red!30, draw=red, area legend] coordinates {({1},0.019) };
          \label{B}
          \addplot[fill=blue!30, draw=blue, area legend] coordinates {({1},0.08) };
          \label{R}
          \addplot[fill=teal!30, draw=teal, area legend] coordinates {({1},0.41) };
          \label{T}
    \end{axis}
    }

    \scriptsize{
    \begin{axis}[
      at={(10.52em,0)},
      ymajorgrids,
      grid style=dashed,
      legend style={at={(0.02,0.60)}, anchor=south west},
      ybar=5pt,
      enlarge x limits=0.5,
      xtick align=inside,
      height=.24\textwidth,
      width=.19\textwidth,
      bar width=0.8em,
      xlabel={(b) \ IE of self-attn\\weight},
      ylabel={Information entropy},
      symbolic x coords={{1}},
      xtick=data,
      nodes near coords align={vertical},
      ymin=2,
      ymax=3.5,
      ytick={2.3,2.7,3.2},
      xticklabels={},
      ylabel style={yshift=-3em},xlabel style={yshift=1.8em,align=center},
      yticklabel style={/pgf/number format/fixed,/pgf/number format/fixed zerofill,/pgf/number format/precision=2,rotate=90},
      ]
          \addplot[fill=red!30, draw=red,area legend] coordinates {({1},3.159) };

          \addplot[fill=blue!30, draw=blue,area legend] coordinates {({1},3.29) };

          \addplot[fill=teal!30, draw=teal,area legend] coordinates {({1},2.64) };

    \end{axis}
    }

    \scriptsize{
    \begin{axis}[
      at={(20.74em,0)},
      ymajorgrids,
      grid style=dashed,
      legend style={at={(0.02,0.60)}, anchor=south west},
      ybar=5pt,
      enlarge x limits=0.5,
      xtick align=inside,
      height=.24\textwidth,
      width=.19\textwidth,
      bar width=0.8em,
      xlabel={(c) \ IE of cross-attn\\weight},
      ylabel={Information entropy},
      symbolic x coords={{1}},
      xtick=data,
      nodes near coords align={vertical},
      ymin=0.06,
      ymax=0.17,
      ytick={0.08,0.11,0.15},
      xticklabels={},
      ylabel style={yshift=-3em},xlabel style={yshift=1.8em,align=center},
      yticklabel style={/pgf/number format/fixed,/pgf/number format/fixed zerofill,/pgf/number format/precision=2,rotate=90},
      ]
          \addplot[fill=red!30, draw=red,area legend] coordinates {({1},0.1017) };

          \addplot[fill=blue!30, draw=blue,area legend] coordinates {({1},0.1465)};

          \addplot[fill=teal!30, draw=teal,area legend] coordinates {({1},0.1460) };

    \end{axis}
    }

\end{tikzpicture}
\caption{Effects of the textual adapter.}\label{IE}
\end{figure}
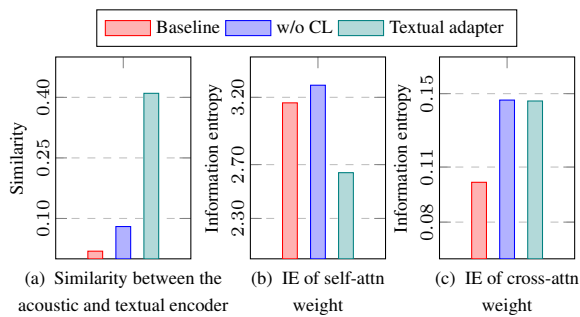

Figure \ref{IE} (a) shows our textual adapter can significantly mitigate the gap between the acoustic and textual model. Figures \ref{IE} (b) and (c) show the average information entropy (IE) of attention weights. Note the IE also has noise information. The IE of the textual adapter is much lower due to the inattention of noise. The IE of w/o CL is the highest, because its adapter  learns more semantic information but fails to remove the noise. Figure \ref{IE} (c) shows the usage of adapter can boost the decoder to extract more information. It also proves that our gains are mainly from the improvement of encoder since the IEs of the two adapters are similar.

\subsection{Sample Efficiency}
We study how different systems behave under different sized speech translation data. To do this, we scale the training data by about 6 times each time. Table \ref{fewshot} shows that our model obtains a good speech translation result by only 10-hour labeled data, which is better than vanilla Transformer \cite{wang-etal-2020-fairseq} learned on 65-hour labeled data. The improvement is still significant when more data is used. This phenomenon shows that the MSP-ST method can build a decent system under the condition of insufficient ST data.

\begin{figure}[t]
  \setlength{\abovecaptionskip}{0.1cm}
  \centering
  \begin{tikzpicture}
    \scriptsize{
    \begin{axis}[
      at={(0,0)},
      ymajorgrids,
      xmajorgrids,
      grid style=dashed,
      legend style={at={(0.02,0.68)}, anchor=south west},
      legend cell align={left},
      ybar,
      xtick align=inside,
      height=.25\textwidth,
      width=.27\textwidth,
      bar width=0.5em,
      xlabel={(a) \ Cross modal alignment},
      ylabel={Similarity},
      symbolic x coords={{1}, {2}, {3},{4},{5},{6}},
      xtick=data,
      nodes near coords align={vertical},
      ymin=-0.05,
      ymax=0.24,
      legend entries={Baseline,Alignment adapter},
      enlarge x limits=0.1,
      xticklabels={the,and,we,or,quickly,strong},
      x tick label style={
            rotate=45,
            anchor=east,
            xshift=0.1cm,
            yshift=-0.15cm
        },
      ylabel style={yshift=-3em},xlabel style={yshift=-0.8em},
      yticklabel style={/pgf/number format/fixed,/pgf/number format/fixed zerofill,/pgf/number format/precision=1,rotate=90},
      ]
      \addplot[fill=red!30, draw=red,area legend] coordinates {({1},-0.0047) ({2},-0.0041) ({3},0.0151) ({4},-0.0294) ({5},0.0090) ({6},-0.0122) };

      \addplot[fill=blue!30, draw=blue,area legend] coordinates {({1},0.1156) ({2},0.1012) ({3},0.1428) ({4},0.1006) ({5},0.1061) ({6},0.0869) };

    \end{axis}
    }

    \scriptsize{
    \begin{axis}[
      at={(15.6em,0)},
      ymajorgrids,
      xmajorgrids,
      grid style=dashed,
      legend style={at={(0.02,0.68)}, anchor=south west},
      legend cell align={left},
      ybar,
      xtick align=inside,
      height=.25\textwidth,
      width=.27\textwidth,
      bar width=0.5em,
      xlabel={(b) \ Cross lingual alignment},
      ylabel={Similarity},
      symbolic x coords={{1},{2},{3},{4},{5},{6}},
      xtick=data,
      nodes near coords align={vertical},
      ymin=-0.05,
      ymax=0.24,
      xticklabels={the,and,we,or,quickly,strong},
      x tick label style={
            rotate=45,
            anchor=east,
            xshift=0.1cm,
            yshift=-0.15cm
        },
      legend entries={Basline, Alignment adapter},
      enlarge x limits=0.1,
      ylabel style={yshift=-3em},xlabel style={yshift=-0.8em},
      yticklabel style={/pgf/number format/fixed,/pgf/number format/fixed zerofill,/pgf/number format/precision=1,rotate=90},
      ]
       \addplot[fill=red!30, draw=red,area legend] coordinates {({1},-0.0030) ({2},-0.0152) ({3},-0.0090) ({4},-0.0234) ({5},-0.0011) ({6},-0.0167) };

       \addplot[fill=blue!30, draw=blue,area legend] coordinates {({1},0.0560) ({2},0.0825) ({3},0.0605) ({4},0.0777) ({5},0.1165) ({6},0.0704) };

    \end{axis}
    }

\end{tikzpicture}
\caption{Similarity of cross-modal and cross-lingual between acoustic and textual modal.}\label{similarity_alignment}
\end{figure}
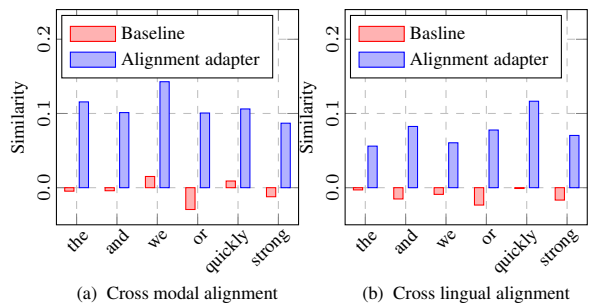

\subsection{Parameter Efficiency}
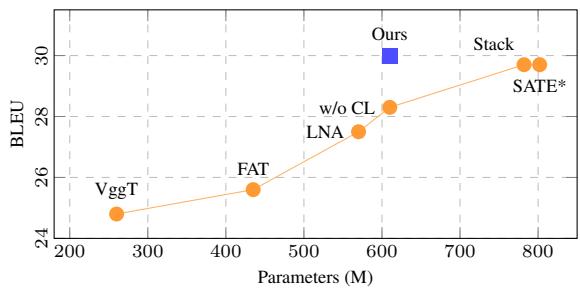
\begin{figure}[t]
  \setlength{\abovecaptionskip}{0.1cm}
  \centering
  \begin{tikzpicture}
  \scriptsize{
    \begin{axis}
    [
	 at={(0em,0)},
      ymajorgrids,
      xmajorgrids,
      grid style=dashed,
      width=.48\textwidth,
      height=.26
      \textwidth,
      legend style={at={(0.12,0.635)}, anchor=south west},
      xlabel={\scriptsize{Parameters (M)}},
      ylabel={\scriptsize{BLEU}},
      ylabel style={yshift=-3.0em},xlabel style={yshift=0.8em},
      yticklabel style={rotate=90},
      ymin=24,ymax=31.5,
      xmin=180,xmax=850,
      xtick={},
      legend style={at={(0.7, 0.1)}, legend plot pos=right,font={\tiny},cells={anchor=west}}
      ]

      \addplot[orange!60] coordinates {
      (260,24.8)
      (435,25.6)
      (570,27.5)
      (610,28.3)
      (782,29.7)
      (802,29.7)
      };
      \node[label={90:{VggT}},circle,fill=orange!80,inner sep=2pt] at (axis cs:260,24.8) {};
      \node[label={FAT},circle,fill=orange!80,inner sep=2pt] at (axis cs:435,25.6) {};
      \node[label={180:{LNA}},circle,fill=orange!80,inner sep=2pt] at (axis cs:570,27.5) {};
      \node[label={180:{w/o CL}},circle,fill=orange!80,inner sep=2pt] at (axis cs:610,28.3) {};
      \node[label={110:{Stack}},circle,fill=orange!80,inner sep=2pt] at (axis cs:782,29.7) {};
      \node[label={270:{SATE*}},circle,fill=orange!80,inner sep=2pt] at (axis cs:802,29.7) {};

      \node[label={Ours},fill=blue!70,inner sep=3pt] at (axis cs:610,30.0) {};


    \end{axis}}
    \end{tikzpicture}
    \caption{Efficiency of parameters. The \textit{Stack} means stacking the acoustic model and the SIDAE. ``SATE*'' represents that we reproduce the SATE method. 
    Note the acoustic model contains the alignment adapter.
    }\label{Efficiency}
\end{figure}

Using pre-training models in general leads to a significant increase of model parameters. To evaluate the efficiency of model size, we compare the performance and parameter number of different methods in Figure \ref{Efficiency}. The upper left of the figure means a higher efficiency. We see that our method is efficient: it achieves the best BLEU score with a slight increase of the number of the parameters. On the performance side, the models which stack the pre-trained acoustic and the whole SIDAE model also achieve comparable performance with our MSP-ST system. But our model is more parameter efficient for dropping the pretrained textual encoder at the ST tuning stage. The w/o CL model with the same parameters performs much worse compared with MSP-ST due to the lack of the denoising ability. We implement the SATE method by using the same data and model size. The SATE does not outperform the \textit{Stack} and MSP-ST methods,
we conjecture that the alignment adapter has bridged the gap between two pretrained models.


\subsection{Impact of data}
We remove the additional unlabeled data and implement the MSP-ST method in Table \ref{Ablaiton_data}. To make a more fair comparison with other work, we use a small-scale parameter named MSP-ST-S. Note both the two methods do not use the knowledge distillation and SpecAugment\cite{Park_2019}. It shows that our method is effective without unlabeled data, and achieves comparable performance with our baseline which trains by massive unlabeled data. This shows previous work fails to use the unlabeled data effectively.


\begin{table}[t]
    \centering
    \small
    \setlength{\tabcolsep}{3mm}{
    \begin{tabular}{lccccr}
    \toprule
      &\makecell[c]{MuST-C En-De} & \#Params \\
    \midrule
    MSP-ST-S & 27.6 &138M  \\
    SATE& 27.3 &- \\
    \bottomrule
    \end{tabular}
    }
    \caption{Comparison of two methods without using unlabeled data.}
    \label{Ablaiton_data}
\end{table}

\section{Related Work}
One aspect of ST where there has already been substantial success is the cascaded model of ASR and MT \cite{758176,matusov2005integration,mathias2006statistical}. An obvious next step is towards end-to-end ST but initial work attempting to develop fully end-to-end systems on limited labeled data has met with much less success in competing the cascaded counterpart \cite{berard2016listen}. This motivates an active line of research on introducing unlabeled data into ST. A straightforward method is to train ST models by additional ASR and/or MT supervision signals, as in multi-task learning \cite{anastasopoulos-chiang-2018-tied,le-etal-2020-dual,9414159,9415058,han-etal-2021-learning}. Similar ideas can be found in other related work, including pseudo data generation \cite{pino2019harnessing,pino20_interspeech}, meta-learning \cite{indurthi2020end}, knowledge distillation \cite{liu19d_interspeech, jia2019leveraging} and curriculum learning \cite{wang-etal-2020-curriculum}.

For stronger results, a number of recent studies focus on pre-training components of ST systems and fine-tuning them on labeled ST data \cite{weiss2017sequence,8461690,pmlr-v139-zheng21a,li-etal-2021-multilingual}. Although these systems are of different model designs, researchers are aware that simply incorporating pre-trained ASR and MT models into ST does not work \cite{wang2020bridging, xu-etal-2021-stacked}, because there is a great length difference between acoustic sequence and word sequence, and the two models have different scopes of encoding, i.e., the ASR model is locally attentive, while the MT model, which represents sentence semantics, is more globally attentive.

Several research groups address this by using an additional encoding network to adapt acoustic encoding to text-like encoding \cite{dong2021listen,tang-etal-2021-improving,li-etal-2021-multilingual, xu-etal-2021-stacked}. \citet{Zhang2022SpeechUTBS} propose the hidden-unit pre-training  and \citet{ye-etal-2022-cross} use the contrastive learning to mitigate the inconsistent representation between the speech and text.
Here we explicitly design a trainable text encoder to link ASR and MT pre-training. Perhaps the most related work to what is doing here is \citet{li-etal-2021-multilingual}'s work. Their system benefits from encoder-decoder pre-training by a text-based BART-like method, but the text encoder is discarded when they train the ST encoder. In this work we find that the involvement of the text encoder in the entire pre-training pipeline is critical to achieve the state-of-the-art performance. We thus share the text encoder in both ASR-based and MT-based pre-training.

Also, it is well-known that silent moments often appear in the acoustic model output but not in MT data. This is in general addressed by either down-sampling the output sequence of the acoustic model \cite{dong2021consecutive,liu2020bridging} or converting the source text to the imitation of the acoustic output by CTC paths \cite{wang2020bridging}. Here we instead develop a simple denoising method to enhance the ability of the text encoder in dealing with normal and noisy sentences.

The adapter has been introduced to bridge the modality gap \cite{xu-etal-2021-stacked} or length gap between acoustic and textual model \cite{li-etal-2021-multilingual}. We further consider the robust modeling and the inconsistency of local and global preferences between the two models. And our adapters avoid the use of the whole textual encoder.

\section{Conclusions}
We explore methods to pre-train all the components of an ST model on labeled and unlabeled speech and text data. To improve the ST encoder, we develop an alignment adapter and textual adapter. Then, we use a text-based pre-trained encoder to bridge the acoustic encoding and text encoding. In addition, we use contrastive training and denoising training to resist the influence of silent moments in speech. Experiments show our system achieves SOTA results on the MuST-C En-De, En-Fr and LibriSpeech En-Fr tasks.

\section{Acknowledgments}
This work was supported in part by the National Science Foundation of China (Nos. 62276056, 61732005 and  61876035), the China HTRD Center Project (No. 2020AAA0107904), Yunnan Provincial Major Science and Technology Special Plan Projects (Nos. 202002AD080001 and 202103AA080015), National Frontiers Science Center for Industrial Intelligence and Systems Optimization (Northeastern University, China. No. B16009) and the Fundamental Research Funds for the Central Universities. The authors would like to thank anonymous reviewers for their comments.

\bibliography{aaai22}

\begin{thebibliography}{50}
\providecommand{\natexlab}[1]{#1}

\bibitem[{Anastasopoulos and Chiang(2018)}]{anastasopoulos-chiang-2018-tied}
Anastasopoulos, A.; and Chiang, D. 2018.
\newblock Tied Multitask Learning for Neural Speech Translation.
\newblock In \emph{Proc. of NAACL}.

\bibitem[{{Baevski} et~al.(2020){Baevski}, {Zhou}, {Mohamed}, and
  {Auli}}]{baevski2020wav2vec}
{Baevski}, A.; {Zhou}, Y.; {Mohamed}, A.; and {Auli}, M. 2020.
\newblock wav2vec 2.0: A Framework for Self-Supervised Learning of Speech
  Representations.
\newblock In \emph{Proc. of NeurIPS}.

\bibitem[{Bapna and Firat(2019)}]{bapna-firat-2019-simple}
Bapna, A.; and Firat, O. 2019.
\newblock Simple, Scalable Adaptation for Neural Machine Translation.
\newblock In \emph{Proc. of EMNLP}.

\bibitem[{B{\'e}rard et~al.(2016)B{\'e}rard, Pietquin, Servan, and
  Besacier}]{berard2016listen}
B{\'e}rard, A.; Pietquin, O.; Servan, C.; and Besacier, L. 2016.
\newblock Listen and translate: A proof of concept for end-to-end
  speech-to-text translation.
\newblock \emph{arXiv preprint arXiv:1612.01744}.

\bibitem[{Bérard et~al.(2018)Bérard, Besacier, Kocabiyikoglu, and
  Pietquin}]{8461690}
Bérard, A.; Besacier, L.; Kocabiyikoglu, A.~C.; and Pietquin, O. 2018.
\newblock End-to-End Automatic Speech Translation of Audiobooks.
\newblock In \emph{Proc. of ICASSP}.

\bibitem[{Chen et~al.(2021)Chen, Wu, Chen, Wu, Li, Yoshioka, Wang, Liu, and
  Zhou}]{9413423}
Chen, S.; Wu, Y.; Chen, Z.; Wu, J.; Li, J.; Yoshioka, T.; Wang, C.; Liu, S.;
  and Zhou, M. 2021.
\newblock Continuous Speech Separation with Conformer.
\newblock In \emph{Proc. of ICASSP}.

\bibitem[{Di~Gangi et~al.(2019)Di~Gangi, Cattoni, Bentivogli, Negri, and
  Turchi}]{di-gangi-etal-2019-must}
Di~Gangi, M.~A.; Cattoni, R.; Bentivogli, L.; Negri, M.; and Turchi, M. 2019.
\newblock {M}u{ST}-{C}: a {M}ultilingual {S}peech {T}ranslation {C}orpus.
\newblock In \emph{Proc. of NAACL}.

\bibitem[{Dong et~al.(2021{\natexlab{a}})Dong, Wang, Zhou, Xu, Xu, and
  Li}]{dong2021consecutive}
Dong, Q.; Wang, M.; Zhou, H.; Xu, S.; Xu, B.; and Li, L. 2021{\natexlab{a}}.
\newblock Consecutive decoding for speech-to-text translation.
\newblock In \emph{Proc. of AAAI}.

\bibitem[{Dong et~al.(2021{\natexlab{b}})Dong, Ye, Wang, Zhou, Xu, Xu, and
  Li}]{dong2021listen}
Dong, Q.; Ye, R.; Wang, M.; Zhou, H.; Xu, S.; Xu, B.; and Li, L.
  2021{\natexlab{b}}.
\newblock “Listen, Understand and Translate”: Triple Supervision Decouples
  End-to-end Speech-to-text Translation.
\newblock In \emph{Proc. of AAAI}.

\bibitem[{G{\'a}llego et~al.(2021)G{\'a}llego, Tsiamas, Escolano, Fonollosa,
  and Costa-juss{\`a}}]{gallego2021end}
G{\'a}llego, G.~I.; Tsiamas, I.; Escolano, C.; Fonollosa, J.~A.; and
  Costa-juss{\`a}, M.~R. 2021.
\newblock End-to-End Speech Translation with Pre-trained Models and Adapters:
  UPC at IWSLT 2021.
\newblock In \emph{Proceedings of the 18th International Conference on Spoken
  Language Translation (IWSLT 2021)}.

\bibitem[{Graves et~al.(2006)Graves, Fern{\'a}ndez, Gomez, and
  Schmidhuber}]{graves2006connectionist}
Graves, A.; Fern{\'a}ndez, S.; Gomez, F.; and Schmidhuber, J. 2006.
\newblock Connectionist temporal classification: labelling unsegmented sequence
  data with recurrent neural networks.
\newblock In \emph{Proc. of ICML}.

\bibitem[{Han et~al.(2021)Han, Wang, Ji, and Li}]{han-etal-2021-learning}
Han, C.; Wang, M.; Ji, H.; and Li, L. 2021.
\newblock Learning Shared Semantic Space for Speech-to-Text Translation.
\newblock In \emph{Proc. of ACL Findings}.

\bibitem[{Hsu et~al.(2021)Hsu, Bolte, Tsai, Lakhotia, Salakhutdinov, and
  Mohamed}]{9585401}
Hsu, W.-N.; Bolte, B.; Tsai, Y.-H.~H.; Lakhotia, K.; Salakhutdinov, R.; and
  Mohamed, A. 2021.
\newblock HuBERT: Self-Supervised Speech Representation Learning by Masked
  Prediction of Hidden Units.
\newblock \emph{IEEE/ACM Transactions on Audio, Speech, and Language
  Processing}.

\bibitem[{Indurthi et~al.(2020)Indurthi, Han, Lakumarapu, Lee, Chung, Kim, and
  Kim}]{indurthi2020end}
Indurthi, S.; Han, H.; Lakumarapu, N.~K.; Lee, B.; Chung, I.; Kim, S.; and Kim,
  C. 2020.
\newblock End-end speech-to-text translation with modality agnostic
  meta-learning.
\newblock In \emph{Proc. of ICASSP}.

\bibitem[{Indurthi et~al.(2021)Indurthi, Zaidi, Kumar~Lakumarapu, Lee, Han,
  Ahn, Kim, Kim, and Hwang}]{9414703}
Indurthi, S.; Zaidi, M.~A.; Kumar~Lakumarapu, N.; Lee, B.; Han, H.; Ahn, S.;
  Kim, S.; Kim, C.; and Hwang, I. 2021.
\newblock Task Aware Multi-Task Learning for Speech to Text Tasks.
\newblock In \emph{Proc. of ICASSP}.

\bibitem[{Jia et~al.(2019)Jia, Johnson, Macherey, Weiss, Cao, Chiu, Ari,
  Laurenzo, and Wu}]{jia2019leveraging}
Jia, Y.; Johnson, M.; Macherey, W.; Weiss, R.~J.; Cao, Y.; Chiu, C.-C.; Ari,
  N.; Laurenzo, S.; and Wu, Y. 2019.
\newblock Leveraging weakly supervised data to improve end-to-end
  speech-to-text translation.
\newblock In \emph{Proc. of ICASSP}.

\bibitem[{Kahn et~al.(2020)Kahn, Riviere, Zheng, Kharitonov, Xu, Mazar{\'e},
  Karadayi, Liptchinsky, Collobert, Fuegen et~al.}]{kahn2020libri}
Kahn, J.; Riviere, M.; Zheng, W.; Kharitonov, E.; Xu, Q.; Mazar{\'e}, P.-E.;
  Karadayi, J.; Liptchinsky, V.; Collobert, R.; Fuegen, C.; et~al. 2020.
\newblock Libri-light: A benchmark for asr with limited or no supervision.
\newblock In \emph{Proc. of ICASSP}.

\bibitem[{Kocabiyikoglu, Besacier, and
  Kraif(2018)}]{kocabiyikoglu-etal-2018-augmenting}
Kocabiyikoglu, A.~C.; Besacier, L.; and Kraif, O. 2018.
\newblock Augmenting Librispeech with {F}rench Translations: A Multimodal
  Corpus for Direct Speech Translation Evaluation.
\newblock In \emph{Proc. of LREC}.

\bibitem[{Le et~al.(2020)Le, Pino, Wang, Gu, Schwab, and
  Besacier}]{le-etal-2020-dual}
Le, H.; Pino, J.; Wang, C.; Gu, J.; Schwab, D.; and Besacier, L. 2020.
\newblock Dual-decoder Transformer for Joint Automatic Speech Recognition and
  Multilingual Speech Translation.
\newblock In \emph{Proc. of COLING}.

\bibitem[{Lewis et~al.(2020)Lewis, Liu, Goyal, Ghazvininejad, Mohamed, Levy,
  Stoyanov, and Zettlemoyer}]{lewis-etal-2020-bart}
Lewis, M.; Liu, Y.; Goyal, N.; Ghazvininejad, M.; Mohamed, A.; Levy, O.;
  Stoyanov, V.; and Zettlemoyer, L. 2020.
\newblock {BART}: Denoising Sequence-to-Sequence Pre-training for Natural
  Language Generation, Translation, and Comprehension.
\newblock In \emph{Proc. of ACL}.

\bibitem[{Li et~al.(2021)Li, Wang, Tang, Tran, Tang, Pino, Baevski, Conneau,
  and Auli}]{li-etal-2021-multilingual}
Li, X.; Wang, C.; Tang, Y.; Tran, C.; Tang, Y.; Pino, J.; Baevski, A.; Conneau,
  A.; and Auli, M. 2021.
\newblock Multilingual Speech Translation from Efficient Finetuning of
  Pretrained Models.
\newblock In \emph{Proc. of ACL}.

\bibitem[{Liu et~al.(2020{\natexlab{a}})Liu, Gu, Goyal, Li, Edunov,
  Ghazvininejad, Lewis, and Zettlemoyer}]{liu-etal-2020-multilingual-denoising}
Liu, Y.; Gu, J.; Goyal, N.; Li, X.; Edunov, S.; Ghazvininejad, M.; Lewis, M.;
  and Zettlemoyer, L. 2020{\natexlab{a}}.
\newblock Multilingual Denoising Pre-training for Neural Machine Translation.
\newblock \emph{Transactions of the Association for Computational Linguistics}.

\bibitem[{Liu et~al.(2019)Liu, Xiong, Zhang, He, Wu, Wang, and
  Zong}]{liu19d_interspeech}
Liu, Y.; Xiong, H.; Zhang, J.; He, Z.; Wu, H.; Wang, H.; and Zong, C. 2019.
\newblock {End-to-End Speech Translation with Knowledge Distillation}.
\newblock In \emph{Proc. of Interspeech}.

\bibitem[{Liu et~al.(2020{\natexlab{b}})Liu, Zhu, Zhang, and
  Zong}]{liu2020bridging}
Liu, Y.; Zhu, J.; Zhang, J.; and Zong, C. 2020{\natexlab{b}}.
\newblock Bridging the modality gap for speech-to-text translation.
\newblock \emph{arXiv preprint arXiv:2010.14920}.

\bibitem[{Mathias and Byrne(2006)}]{mathias2006statistical}
Mathias, L.; and Byrne, W. 2006.
\newblock Statistical phrase-based speech translation.
\newblock In \emph{2006 IEEE International Conference on Acoustics Speech and
  Signal Processing Proceedings}.

\bibitem[{Matusov, Kanthak, and Ney(2005)}]{matusov2005integration}
Matusov, E.; Kanthak, S.; and Ney, H. 2005.
\newblock On the integration of speech recognition and statistical machine
  translation.
\newblock In \emph{Ninth European Conference on Speech Communication and
  Technology}.

\bibitem[{Ney(1999)}]{758176}
Ney, H. 1999.
\newblock Speech translation: coupling of recognition and translation.
\newblock In \emph{Proc. of ICASSP}.

\bibitem[{Ott et~al.(2019)Ott, Edunov, Baevski, Fan, Gross, Ng, Grangier, and
  Auli}]{ott2019fairseq}
Ott, M.; Edunov, S.; Baevski, A.; Fan, A.; Gross, S.; Ng, N.; Grangier, D.; and
  Auli, M. 2019.
\newblock fairseq: A Fast, Extensible Toolkit for Sequence Modeling.
\newblock In \emph{Proc. of AACL}.

\bibitem[{Panayotov et~al.(2015)Panayotov, Chen, Povey, and
  Khudanpur}]{7178964}
Panayotov, V.; Chen, G.; Povey, D.; and Khudanpur, S. 2015.
\newblock Librispeech: An ASR corpus based on public domain audio books.
\newblock In \emph{Proc. of ICASSP}.

\bibitem[{Park et~al.(2019)Park, Chan, Zhang, Chiu, Zoph, Cubuk, and
  Le}]{Park_2019}
Park, D.~S.; Chan, W.; Zhang, Y.; Chiu, C.-C.; Zoph, B.; Cubuk, E.~D.; and Le,
  Q.~V. 2019.
\newblock {SpecAugment}: A Simple Data Augmentation Method for Automatic Speech
  Recognition.
\newblock In \emph{Proc. of Interspeech}.

\bibitem[{Pino et~al.(2019)Pino, Puzon, Gu, Ma, McCarthy, and
  Gopinath}]{pino2019harnessing}
Pino, J.; Puzon, L.; Gu, J.; Ma, X.; McCarthy, A.~D.; and Gopinath, D. 2019.
\newblock Harnessing indirect training data for end-to-end automatic speech
  translation: Tricks of the trade.
\newblock \emph{arXiv preprint arXiv:1909.06515}.

\bibitem[{Pino et~al.(2020)Pino, Xu, Ma, Dousti, and Tang}]{pino20_interspeech}
Pino, J.; Xu, Q.; Ma, X.; Dousti, M.~J.; and Tang, Y. 2020.
\newblock {Self-Training for End-to-End Speech Translation}.
\newblock In \emph{Proc. Interspeech 2020}, 1476--1480.

\bibitem[{Post(2018)}]{post-2018-call}
Post, M. 2018.
\newblock A Call for Clarity in Reporting {BLEU} Scores.
\newblock In \emph{Proceedings of the Third Conference on Machine Translation:
  Research Papers}.

\bibitem[{Tang et~al.(2022)Tang, Gong, Dong, Wang, Hsu, Gu, Baevski, Li,
  Mohamed, Auli et~al.}]{tang2022unified}
Tang, Y.; Gong, H.; Dong, N.; Wang, C.; Hsu, W.-N.; Gu, J.; Baevski, A.; Li,
  X.; Mohamed, A.; Auli, M.; et~al. 2022.
\newblock Unified Speech-Text Pre-training for Speech Translation and
  Recognition.
\newblock \emph{arXiv preprint arXiv:2204.05409}.

\bibitem[{Tang et~al.(2021{\natexlab{a}})Tang, Pino, Li, Wang, and
  Genzel}]{tang-etal-2021-improving}
Tang, Y.; Pino, J.; Li, X.; Wang, C.; and Genzel, D. 2021{\natexlab{a}}.
\newblock Improving Speech Translation by Understanding and Learning from the
  Auxiliary Text Translation Task.
\newblock In \emph{Proc. of ACL}.

\bibitem[{Tang et~al.(2021{\natexlab{b}})Tang, Pino, Wang, Ma, and
  Genzel}]{9415058}
Tang, Y.; Pino, J.; Wang, C.; Ma, X.; and Genzel, D. 2021{\natexlab{b}}.
\newblock A General Multi-Task Learning Framework to Leverage Text Data for
  Speech to Text Tasks.
\newblock In \emph{Proc. of ICASSP}.

\bibitem[{Vaswani et~al.(2017)Vaswani, Shazeer, Parmar, Uszkoreit, Jones,
  Gomez, Kaiser, and Polosukhin}]{vaswani2017attention}
Vaswani, A.; Shazeer, N.; Parmar, N.; Uszkoreit, J.; Jones, L.; Gomez, A.~N.;
  Kaiser, {\L}.; and Polosukhin, I. 2017.
\newblock Attention is all you need.
\newblock In \emph{Proc. of NeurIPS}.

\bibitem[{Vydana et~al.(2021)Vydana, Karafiát, Zmolikova, Burget, and
  Černocký}]{9414159}
Vydana, H.~K.; Karafiát, M.; Zmolikova, K.; Burget, L.; and Černocký, H.
  2021.
\newblock Jointly Trained Transformers Models for Spoken Language Translation.
\newblock In \emph{Proc. of ICASSP}.

\bibitem[{Wang et~al.(2020{\natexlab{a}})Wang, Tang, Ma, Wu, Okhonko, and
  Pino}]{wang-etal-2020-fairseq}
Wang, C.; Tang, Y.; Ma, X.; Wu, A.; Okhonko, D.; and Pino, J.
  2020{\natexlab{a}}.
\newblock Fairseq {S}2{T}: Fast Speech-to-Text Modeling with Fairseq.
\newblock In \emph{Proc. of AACL}.

\bibitem[{Wang et~al.(2020{\natexlab{b}})Wang, Wu, Liu, Yang, and
  Zhou}]{wang2020bridging}
Wang, C.; Wu, Y.; Liu, S.; Yang, Z.; and Zhou, M. 2020{\natexlab{b}}.
\newblock Bridging the gap between pre-training and fine-tuning for end-to-end
  speech translation.
\newblock In \emph{Proc. of AAAI}.

\bibitem[{Wang et~al.(2020{\natexlab{c}})Wang, Wu, Liu, Zhou, and
  Yang}]{wang-etal-2020-curriculum}
Wang, C.; Wu, Y.; Liu, S.; Zhou, M.; and Yang, Z. 2020{\natexlab{c}}.
\newblock Curriculum Pre-training for End-to-End Speech Translation.
\newblock In \emph{Proc. of ACL}.

\bibitem[{Watanabe et~al.(2017)Watanabe, Hori, Kim, Hershey, and
  Hayashi}]{watanabe2017hybrid}
Watanabe, S.; Hori, T.; Kim, S.; Hershey, J.~R.; and Hayashi, T. 2017.
\newblock Hybrid CTC/attention architecture for end-to-end speech recognition.
\newblock \emph{IEEE Journal of Selected Topics in Signal Processing}.

\bibitem[{Weiss et~al.(2017)Weiss, Chorowski, Jaitly, Wu, and
  Chen}]{weiss2017sequence}
Weiss, R.~J.; Chorowski, J.; Jaitly, N.; Wu, Y.; and Chen, Z. 2017.
\newblock Sequence-to-sequence models can directly translate foreign speech.
\newblock \emph{arXiv preprint arXiv:1703.08581}.

\bibitem[{Xu et~al.(2021)Xu, Hu, Li, Zhang, Huang, Ju, Xiao, and
  Zhu}]{xu-etal-2021-stacked}
Xu, C.; Hu, B.; Li, Y.; Zhang, Y.; Huang, S.; Ju, Q.; Xiao, T.; and Zhu, J.
  2021.
\newblock Stacked Acoustic-and-Textual Encoding: Integrating the Pre-trained
  Models into Speech Translation Encoders.
\newblock In \emph{Proc. of ACL}.

\bibitem[{Ye, Wang, and Li(2021)}]{ye2021end}
Ye, R.; Wang, M.; and Li, L. 2021.
\newblock End-to-end Speech Translation via Cross-modal Progressive Training.
\newblock \emph{arXiv preprint arXiv:2104.10380}.

\bibitem[{Ye, Wang, and Li(2022)}]{ye-etal-2022-cross}
Ye, R.; Wang, M.; and Li, L. 2022.
\newblock Cross-modal Contrastive Learning for Speech Translation.
\newblock In \emph{Proceedings of the 2022 Conference of the North American
  Chapter of the Association for Computational Linguistics: Human Language
  Technologies}, 5099--5113. Seattle, United States: Association for
  Computational Linguistics.

\bibitem[{Zhang et~al.(2020)Zhang, Zhang, Odena, and
  Lee}]{Zhang2020Consistency}
Zhang, H.; Zhang, Z.; Odena, A.; and Lee, H. 2020.
\newblock Consistency Regularization for Generative Adversarial Networks.
\newblock In \emph{Proc. of ICLR}.

\bibitem[{Zhang et~al.(2022{\natexlab{a}})Zhang, Huang, Xu, Liu, Li, Ma, Xiao,
  and Zhu}]{zhang-etal-2022-niutranss}
Zhang, Y.; Huang, C.; Xu, C.; Liu, X.; Li, B.; Ma, A.; Xiao, T.; and Zhu, J.
  2022{\natexlab{a}}.
\newblock The {N}iu{T}rans{'}s Submission to the {IWSLT}22
  {E}nglish-to-{C}hinese Offline Speech Translation Task.
\newblock In \emph{Proceedings of the 19th International Conference on Spoken
  Language Translation (IWSLT 2022)}.

\bibitem[{Zhang et~al.(2022{\natexlab{b}})Zhang, Zhou, Ao, Liu, Dai, Li, and
  Wei}]{Zhang2022SpeechUTBS}
Zhang, Z.-H.; Zhou, L.; Ao, J.; Liu, S.; Dai, L.; Li, J.; and Wei, F.
  2022{\natexlab{b}}.
\newblock SpeechUT: Bridging Speech and Text with Hidden-Unit for
  Encoder-Decoder Based Speech-Text Pre-training.
\newblock \emph{ArXiv}, abs/2210.03730.

\bibitem[{Zheng et~al.(2021)Zheng, Chen, Ma, and Huang}]{pmlr-v139-zheng21a}
Zheng, R.; Chen, J.; Ma, M.; and Huang, L. 2021.
\newblock Fused Acoustic and Text Encoding for Multimodal Bilingual Pretraining
  and Speech Translation.
\newblock In \emph{Proc. of ICML}.

\end{thebibliography}

\clearpage
\appendix
\section{Appendix}
\subsection{A Data}
\paragraph{Unlabeled Data} For speech data, we use the Librilight \footnote{https://github.com/facebookresearch/libri-light} to pre-train the acoustic model. It consists of about 60k hours of unlabelled speech. For text data, we followed \citet{liu-etal-2020-multilingual-denoising}'s work which covers 25 languages monolingual sentences. The data corpus is extracted from the Common Crawl\footnote{https://commoncrawl.org/}.
\paragraph{ASR and MT Data} To convert the pre-trained acoustic model on the English ASR task, we use LibriSpeech corpus\footnote{http://www.openslr.org/11/}. It is a read ASR corpus and consisis of 960 hours speech data. To adapt the DAE model to the MT task, we use Opensubtitle En-De and WMT14 En-Fr datasets respectively. The original Opensubtitle En-De \footnote{https://opus.nlpl.eu/OpenSubtitles2018.php} consists of 22M parallel sentences and we filter the parallel data by a max length ratio 1.5 and a max length of 200. The final data size is 18M for En-De translation. For En-Fr translation, we extract 10M sentence pairs from the WMT14 En-Fr data, following \citet{xu-etal-2021-stacked}'s work. 
\paragraph{ST Data} The MuST-C corpus is a multilingual speech translation corpus extracted from TED talks. The size of speech translation data is 400 hours (230K utterances) for the En-De task and 484 hours (270K utterances) for the En-Fr task. For the LibriSpeech En-Fr task, the size of the training set is 100 hours (44K utterances). We remove the utterances of more than 3,000 frames in all the experiments.

\subsection{B Training Details}
\paragraph{Unlabeled data pretraining} For pre-training of unlabeled speech data, we use the open-source wav2vec2 model which is not fine-tuned on the ASR task. For the DAE model, we also utilize the open-source mBART.CC25 model.

\paragraph{MT training stage} For other hyper-parameters to pre-train the SIDAE, we follow the settings in \citet{liu-etal-2020-multilingual-denoising}'s work.
We stop training until the perplexity converges on the valid set.
\paragraph{ASR training stage}
For the alignment adapter, the size of the convolution layer $n$ is set to 3, i.e., we use three 1D convolution layers with a stride of 2. It results in 8 times length compression. For each Conformer layer, there are 1,024 hidden states, 16 attention heads and 4,096 FFN hidden states. For the textual adapter, the configurations of the Conformer layer are the same as the alignment adapter.  
The other settings such as learning rate, training batch and steps are the same as in \cite{baevski2020wav2vec}. 

\paragraph{ST fine-tuning stage} For fine-tuning on the ST task, we use the Adam optimizer with $\beta_{1}=0.9$ and $\beta_{2}=0.98$. Also, we use Dropout ($p=0.1$) and label smoothing ($p=0.1$) for robust training. We early stop the training if the last five checkpoints do not improve and the max training epoch is 20. The number of update parameters is about 198M during the fine-tuning stage. We pre-train our model on the ASR and MT tasks on 8 Nvidia Tesla-V100 GPUs. We fine-tune on the ST task using 4 GPUS with a max token number of 10,000. We use speech as input for our pre-trained model. For Transformer without pre-training, the input speech is represented as 80D log mel-filterbank coefficients that are computed every 10ms with a 25ms window. 

\subsection{C MSP-ST-S Training Details}
The MSP-ST-S is a small model without using unlabeled data. The size of the shared vocab is 10,000, and we use the same vocab during the three training stages. We use the wav2vec 2.0 small model as the acoustic encoder. It has twelve Transformer layers with a 768 hidden size and seven convolution layers. The alignment adapter is a one Conformer layer and the textual adapter is a one Transformer layer. The pre-training machine translation is consist of a twelve-layers encoder and a six-layers decoder. 

\end{document}